\documentclass[showpacs,twocolumn,showkeys,amsmath,amssymb,pra,nofootinbib,subscriptaddress,longbibliography]{revtex4-1}

\usepackage{amsmath,amssymb,amsfonts}
\usepackage{graphicx}
\usepackage{epstopdf}
\usepackage{txfonts}
\usepackage{subfigure}

\usepackage{color}

\usepackage[unicode]{hyperref}
\hypersetup{
   unicode=true,          
   a4paper=true,
   plainpages=false,
   pdftitle={Title of PDF},    
   pdfauthor={Author of PDF},     
   pdfsubject={Subject of PDF},   
   colorlinks=true,       
   citecolor=blue,        
}

\begin{document}

\title{Superballistic center-of-mass motion in one-dimensional
  attractive Bose gases: Decoherence-induced Gaussian random walks in velocity space}

\author{Christoph Weiss}
\email{christoph.weiss@durham.ac.uk}

\author{Simon L. Cornish}

\author{Simon A. Gardiner }

\affiliation{Joint Quantum Centre (JQC) Durham--Newcastle, Department of Physics, Durham University, Durham DH1 3LE, United Kingdom}

\author{Heinz-Peter Breuer}

\affiliation{Physikalisches Institut, Universit\"at Freiburg, Hermann-Herder-Stra{\ss}e 3, D-79104 Freiburg, Germany}

\date{15 December 2015}

 \begin{abstract}
We show that the spreading of the center-of-mass density of ultracold attractively
interacting bosons can become superballistic in the presence of
decoherence, via single-, two- and/or three-body losses. In the limit
of weak decoherence, we analytically solve the numerical model
introduced in [\textit{Phys.\ Rev.\ A}\/ \textbf{91}, 063616
(2015)]. The analytical predictions allow us to identify experimentally accessible parameter
regimes for which we predict superballistic spreading of the center-of-mass density.  
Ultracold attractive Bose gases form weakly bound molecules;
quantum matter-wave bright solitons. Our computer-simulations combine
ideas from classical field methods (``truncated Wigner'') and piecewise deterministic stochastic
processes. While the truncated Wigner approach to use an average
over classical paths as a substitute for a quantum superposition is often
an uncontrolled approximation, here it predicts the exact root-mean-square width when modeling an expanding
Gaussian wave packet. In the superballistic regime, the leading-order of the spreading of the center-of-mass density can thus
be modeled as a quantum superposition of classical Gaussian random walks in velocity space.
\end{abstract}


\maketitle

\section{Introduction}

Superballistic motion (motion with increasing velocities) has been investigated in the context of random walks with random velocities~\cite{ZaburdaevEtAl2008}, driven magnetic turbulence~\cite{ZimbardoEtAl2000}, atom-photon interactions in cavity QED~\cite{PrantsEtAl2002} and nonergodic noise \cite{LiuBao2014}. In quantum systems, time-dependent random potentials have been demonstrated
to cause superballistic transport~\cite{LeviEtAl2012}. Superballistic transport
was predicted theoretically in the dynamics of wave-packet spreading in a tight-binding
lattice junction \cite{HufnagelEtAl2001,ZhangEtAl2012} and observed experimentally in a hybrid
photonic lattice setup~\cite{StutzerEtAl2013}. For a relativistic kicked-rotor system, superballistic transport occurs both in the classical and quantum  regime~\cite{ZhaoEtAl2014}.

The present paper provides an analytical solution of the numerical model for the spreading of the center-of-mass density of a quantum bright soliton under the influence of decoherence via particle losses introduced in Ref.~\cite{WeissEtAl2015}. The analytic approach presented here is valid in the limit that few particles (compared to the total number of particles) are lost. We use this approach to identify experimentally realistic parameters for which we predict that \textit{superballistic}\/ spreading of the center-of-mass density  can be observed experimentally.

 Bright solitons can be
experimentally generated from attractively interacting ultracold
atomic
gases~\cite{KhaykovichEtAl2002,StreckerEtAl2002,CornishEtAl2006,MarchantEtAl2013,MedleyEtAl2014,McDonaldEtAl2014,NguyenEtAl2014,MarchantEtAl2016,EverittEtAl2015}; 
on the mean-field level, via the Gross-Pitaevskii equation (GPE),
these matter-wave bright solitons are nonspreading solutions of a
nonlinear
equation~\cite{PethickSmith2008,BaizakovEtAl2002,AlKhawajaEtAl2002,HaiEtAl2004,MartinRuostekoski2012,CuevasEtAl2013,PoloAhufinger2013,SunEtAl2014,HelmEtAl2015,DunjkoOlshanii2015}. Many-particle quantum descriptions of solitons can be found in Refs.~\cite{LaiHaus1989,DrummondEtAl1993,CarrBrand2004,MishmashCarr2009,StreltsovEtAl2011,FogartyEtAl2013,DelandeEtAl2013,GertjerenkenEtAl2013,BarbieroEtAl2014,DelandeSacha2014,KronkeSchmelcher2014,GertjerenkenKevrekidis2015}.

Beyond enabling us to predict parameters of superballistic spreading of the center-of-mass density, the
analytical solution presented in the present paper of our numerical
model~\cite{WeissEtAl2015} also allows us to quantitatively predict
the timescale on which the transition from \textit{short-time} diffusive to
\textit{long-time} ballistic behavior observed numerically in
Ref.~\cite{WeissEtAl2015}  takes place.\footnote{Models that behave either
  ballistically or diffusively depending on the choice of parameters
  can be found in Ref.~\cite{SteinigewegEtAl2007}.} This behavior is
the opposite of free Brownian
motion~\cite{GrabertEtAl1988,JungHanggi1991,LukicEtAl2005,KoepplEtAl2006}
(cf.~\cite{HanggiEtAl2009,DierlEtAl2014}) which exhibits the generic short-time-scale ballistic and long-time-scale diffusive behavior; for anomalous Brownian motion see~\cite{TurivEtAl2015}. Our model is complementary to previous research both on quantum Brownian motion~\cite{GrabertEtAl1988,FisherZwerger1985} and anomalous diffusion~\cite{Metzler2000} as well as quantum random walks with or without decoherence~\cite{DurEtAl2002,KarskiEtAl2009,AgliariEtAl2010}.

The paper is organized as follows:
Section~\ref{sec:model} introduces models to describe the spreading of the center-of-mass density of bright solitons in attractively interacting Bose gases in the absence of decoherence. In Sec.~\ref{sec:Dec} we extend the model for decoherence-induced spreading of the center-of-mass density of Ref.~\cite{WeissEtAl2015} to include single- and two-particle losses in addition to the dominant three-particle losses. The agreement between analytical and numerical calculations is demonstrated in Sec.~\ref{sec:Results}.  For experimentally accessible parameters (for both \textsuperscript{7}Li and \textsuperscript{85}Rb) we predict superballistic spreading of the center-of-mass density analytically and observe it numerically. The paper ends with conclusions and outlook in Sec.~\ref{sec:Conclusion}.

\section{\label{sec:model}Modeling {spreading of the center-of-mass density}  in the absence of decoherence}

\subsection{\label{sub:model}Overview of Sec.~\ref{sec:model}}

As in Ref.~\cite{WeissEtAl2015}, we consider the physical situation
that the ultracold attractively interacting Bose gas moves in a
quasi-one-dimensional waveguide. An initial weak harmonic trap in the
direction of the waveguide is switched off at $t=0$. For the
definition of ``weak'' we start with the mean-field description of
matter-wave bright solitons (Sec.~\ref{sub:Mean}). While the
center-of-mass wave function of a quantum bright soliton spreads
(Sec.~\ref{sub:Center}), this does not affect the particle density
measured in a single measurement
(Sec.~\ref{sub:Single}). The truncated Wigner
  approximation
is particular suitable to model the spreading of a Gaussian wave packet as it agrees with the exact result (Sec.~\ref{sub:Truncated}).

\subsection{\label{sub:Mean}Mean-field approach via the Gross-Pitaevskii equation}

Often, important aspects of bright solitons can be understood by the one-dimensional Gross-Pitaevskii equation (GPE)~\cite{PethickSmith2008}
\begin{align}
\label{eq:GPE}
i\hbar \frac{\partial}{\partial t}\varphi = -\frac{\hbar^2}{2m}\frac{\partial^2}{\partial x^2}
\varphi +
 \frac {m\omega^2x^2}2\varphi
+g_{1 \rm D}(N-1)|\varphi|^2 \varphi,
\end{align}
where $m$ is the mass of the particles and $\omega$ the angular
frequency of the harmonic trap. The (attractive) interaction
\begin{align}
\label{eq:g1d}
g_{\rm 1D} &=2\hbar
\omega_{\perp}a \\ &<0 \nonumber
\end{align}
 is proportional to the
\textit{s}-wave scattering
length $a$ and the perpendicular angular trapping-frequency,
$\omega_{\perp}$~\cite{Olshanii1998}. 

For attractive interactions ($g_{\rm 1D}<0$) and weak harmonic trapping, Eq.~(\ref{eq:GPE}) has bright-soliton
solutions with single-particle densities $\varrho \equiv |\varphi|^2$~\cite{PethickSmith2008}:
\begin{equation}
\label{eq:singlesoliton}
\varrho(x)= \frac{1}{4\xi_N\left\{\cosh[x/(2\xi_N)]\right\}^2},
\end{equation}
where the soliton length is given by\footnote{This result coincides~\cite{CalogeroDegasperis1975,CastinHerzog2001} with the soliton size derived from the Lieb-Linger model~\cite{LiebLiniger1963} with attractive interactions, appendix~\ref{app:LiebLiniger}.}
\begin{equation}
\label{eq:solitonlength}
 \xi_N \equiv \frac{\hbar^2}{m\left|g_{\rm 1D}\right|(N-1)}.
\end{equation}

If we open a sufficiently weak, that is
  $\xi_N\ll\sqrt{\hbar/(m\omega)}$, initial harmonic trap at $t=0$,
this does not lead to excited atoms as long as the length scale of the
trap is large compared to the soliton length. This has been shown on
the mean-field level in Ref.~\cite{Castin2009} (for a many-particle
version cf.\ Ref.~\cite{HoldawayEtAl2012}). On the GPE-level, opening
a sufficiently weak trap does not lead to any dynamics at all --- not even for the center of mass.

\subsection{\label{sub:Center} spreading of the center-of-mass density of quantum bright solitons}

Without a trapping potential in the $x$-direction, the direction of
the wave guide, physically realistic $N$-particle models are
translationally invariant in the $x$-direction ($y$- and
$z$-directions are harmonically trapped). In such models, the
center-of-mass eigenfunctions in the direction of the wave guide are
plane waves and the center-of-mass dynamics resembles that of a heavy
single particle. Thus, the center-of-mass dynamics are
 described by the Hamiltonian
\begin{equation}
\label{CoMHam}
 \hat{H} = -\frac{\hbar^2}{2Nm}\frac{\partial^2}{\partial X^2};
\end{equation}
where the center-of-mass coordinate   is given by the average of the positions of all $N$ particles
\begin{equation}
\label{eq:CoMCoord}
 X=\frac 1N\sum_{j=1}^Nx_j.
\end{equation}
Even in the presence of a harmonic potential,
the dynamics of the center of mass of an interacting gas are independent of the interactions, giving
rise to the so-called ``Kohn mode''~\cite{BonitzEtAl2007}. 

If we now open the sufficiently weak initial trap
 described at the end of the previous section~\cite{WeissEtAl2015}, this does not affect the internal degrees of freedom of our many-particle bright soliton. The initial center-of-mass wave function is independent of both the interactions and the approximate modeling of these interactions; its time-dependence is given by~\cite{Fluegge1990}
\begin{align}
\label{eq:wavefunction}
\Psi(X,t) \propto &\left(1+i\frac{\hbar
    t}{2M\Delta X_0^2}\right)^{-1/2}\\
\nonumber &
\times \exp\left(-\frac{X^2-{i2\Delta X_0^2MV_0[X-V_0t]/\hbar}}{{4\Delta X_0^2}\left[1+i\hbar
      t/({2M\Delta X_0^2})\right]}\right),
\end{align}
where $X$ is the  center-of-mass coordinate~(\ref{eq:CoMCoord}), $M=Nm$ and $V_0$ the initial velocity. This leads to an
rms width of~\cite{Fluegge1990}
\begin{equation}
\Delta X = \Delta X_0\sqrt{1+\left(\frac{\hbar t}{2M\Delta X_0^2}\right)^2}.
\label{eq:rms}
\end{equation}

\subsection{\label{sub:Single}Single-particle density in the absence of decoherence}

Although the center-of-mass wave function~(\ref{eq:wavefunction}) spreads according to Eq.~(\ref{eq:rms}),
 a single
measurement of the atomic density via scattering light off the soliton (cf.~\cite{KhaykovichEtAl2002}) still yields the density profile of the soliton~(\ref{eq:singlesoliton}), expected both on the mean-field
(GPE) level and on the $N$-particle quantum level for
vanishing width of the center-of-mass wave
function~\cite{CalogeroDegasperis1975,CastinHerzog2001}. 
Taking into account
harmonic trapping perpendicular to the $x$-axis, one obtains the density~\cite{KhaykovichEtAl2002}
\begin{equation}
\label{eq:density}
\varrho(x,y,z) = \frac N{4\xi_N\left\{\cosh\left[{x}/({2\xi_N})\right]\right\}^2}\frac1{\lambda_{\perp}^2{\pi}}\exp\left(-\frac{y^2+z^2}{\lambda_{\perp}^2}\right),
\end{equation}
where 
\begin{equation}
\label{eq:lambda}
\lambda_{\perp}\equiv\sqrt{\frac{\hbar}{m\omega_{\perp}}}
\end{equation}
 is the
perpendicular harmonic oscillator length; the soliton length $\xi_N$ is given by Eq.~(\ref{eq:solitonlength}).

\subsection{\label{sub:Truncated}Truncated Wigner approximation for  the spreading of the center-of-mass density}

Between
 loss events, the quantum dynamics is known analytically
[Eq.~(\ref{eq:wavefunction})]. Instead of solving the Schr\"odinger equation we use a classical field approach~\cite{WeissEtAl2015}: the truncated Wigner
approximation (TWA)\footnote{The truncated-Wigner approximation~\cite{SinatraEtAl2002} describes quantum systems by averaging over realizations of an
appropriate classical field equation (in this case, the GPE) with initial noise
appropriate to either finite~\cite{BieniasEtAl2011} or zero
temperatures~\cite{MartinRuostekoski2012,GertjerenkenEtAl2013,WeissEtAl2015}. } for the center of mass,
which has been used in Ref.~\cite{GertjerenkenEtAl2013} to qualitatively
emulate quantum behavior on the mean-field level by introducing
classical noise mimicking the quantum uncertainties in both position
and momentum of the center of mass. For an expanding Gaussian
wave-packet, the agreement of TWA for the center of mass with full
quantum predictions is even quantitative~\cite{WeissEtAl2015}. Both the mean position and
the variance calculated via the TWA for the center of mass are identical to the quantum mechanical result. In order to make both results identical, Gaussian noise has to be added independently to
both position $X_0\rightarrow X= X_0 + \delta X_0$ and velocity
$V_0\rightarrow V=V_0+ \delta V_0$ with $\langle \delta X_0\rangle =0$ and
$\langle \delta V_0\rangle =0$ and rms fluctuations
$\sigma_X = \Delta X_0$. The rms for the velocity is given
by the
minimal uncertainty relation
\begin{equation}
\sigma_V=\frac{\hbar}{2 M \sigma_X}.
\end{equation}

 The mean position $\overline{x(t)}=\overline{X_0+V_0t}$ is thus
 identical to the quantum mechanical result; the root-mean-square
 fluctuations $\Delta x = \sqrt{(\Delta X_0)^2+(\Delta V_0)^2t^2}$
 coincide with the quantum mechanical equation~(\ref{eq:rms}). Thus,
 in the absence of both the trap in the axial direction and the
 scattering processes investigated in
 Ref.~\cite{GertjerenkenEtAl2013}, the TWA for the center of mass gives exact results for both the position of the center of mass and the root-mean-square fluctuations of the center of mass for a quantum bright soliton.

To summarize this subsection: As long as there are no quantum interferences, the treatment gives the exact rms fluctuations of the center-of-mass position~\cite{WeissEtAl2015}.

\section{\label{sec:Dec}Decoherence via single- two- and
  three-particle losses}
\subsection{\label{sub:Dec}Overview of Sec.~\ref{sec:Dec}}
We numerically model atom losses (Sec.~\ref{sub:Particle}) via a
stochastic approach using piecewise deterministic
processes~\cite{Davis1993}. For a stochastic implementation of such an
approach to decoherence
see~\cite{DalibardEtAl1992,DumEtAl1992,Breuer2006}; for recent
modeling of open quantum systems in the field of cold atoms, for
example, Ref.~\cite{KronkeEtAl2015} and references
therein. Surprisingly~\cite{WeissEtAl2015}, a classical approach
(Sec.~\ref{sub:Classical}) can be used to describe the quantum
mechanical spreading of the center-of-mass wave function (cf.\ Sec.~\ref{sub:Truncated}).

\subsection{\label{sub:Particle}Particle losses}
In order to model
$n$-particle losses we use density-dependent rate equations~\cite{GrimmEtAl2000}
\begin{equation}
\frac{dN}{dt} = -K_{n} \int d^3 r\, \varrho^{n}(x,y,z),
\end{equation}
where $K_n$ is determined empirically and $\varrho^{n}(x,y,z)$ is given by Eq.~(\ref{eq:density}).

\textit{For three-particle losses, $n=3$,} we have
\begin{align}
\label{eq:dNdtmiddleright}
\frac{dN}{dt} &= -\frac{1}{90\pi^2}K_3\frac{N^3}{\xi_N^2\lambda_{\perp}^4} 
= -\frac1{t_3} (N-1)^2N^3 ,
\end{align}
with
\begin{equation}
\label{eq:t3}
t_3 \equiv \frac{90\pi^2\hbar^4\lambda_{\perp}^4}{m^2g_{\rm 1D}^2K_3}
\end{equation}
 we find, for large $N$ \cite{WeissEtAl2015}
\begin{align}
\label{eq:Ntleftright}
N(t) 
\simeq N_0\left(1 + 4N_0^4\frac {t }{t_3}\right)^{-1/4}.
\end{align}

\textit{For two particle losses,  $n=2$,} we have
\begin{align}
\label{eq:dNdtmiddleright2}
\frac{dN}{dt} &= -\frac 1{12\pi}K_2{\frac {{N}^{2}}{{\xi_N\lambda}^{2}}}
= -\frac1{t_2} (N-1)N^2 ,
\end{align}
with
\begin{equation}
\label{eq:t2}
t_2 \equiv \frac{12\pi\hbar^2\lambda_{\perp}^2}{m|g_{\rm 1D}|K_2}
\end{equation}
for which we obtain~\cite{maple}
\begin{align}
\label{eq:Ntleftright2}
N(t) 
\simeq N_0\left(1 + 2N_0^2\frac {t }{t_2}\right)^{-1/2}.
\end{align}

\textit{For single particle losses, $n=1$,} we have
\begin{equation}
\frac{dN}{dt} =-\frac N{t_1}
\end{equation}
with $t_1 = 1/K_1$ and thus
\begin{equation}
N(t) = N_0\exp\left(-\frac t{t_1}\right).
\end{equation}

Combining all three loss-mechanisms together in one analytical formula is also possible. However, it is of the form ``time as a function of $N$, $t=t(N)$,'' rather than the more usual other way round:
\begin{equation}
\label{eq:dt}
dt \simeq -\frac{dN}{\displaystyle\frac {N}{t_1}+ \frac {N^3}{t_2}+\frac {N^5}{t_3}}
\end{equation}
and thus~\cite{maple}
\begin{align}
\label{eq:Nana}
t(N) \equiv&~ F(N) - F(N_0);\\
\label{eq:FofN}
F(N)\simeq & -t_1\,\ln  \left( N \right) +\frac 1 4\,t_1\,\ln  \left( {N}^{4}t_1\,t_2+{N}^{2}t_1\,t_3+t_3\,t_2
 \right)\nonumber\\& +\frac{\displaystyle\frac 12\,t_3\,{t_1}^{2}\arctan \left( {\displaystyle\frac {2\,{N}^{
2}t_1\,t_2+t_1\,t_3}{\sqrt {-{t_1}^{2}{t_3}^{2}+4\,t_3\,{t_2}^{2}t_1}}} \right) }{
\sqrt {-{t_1}^{2}{t_3}^{2}+4\,t_3\,{t_2}^{2}t_1}}.
\end{align}

A very important time-scale is the time in
 which on average one loss event takes place. This time-scale,
\begin{equation}
\label{eq:timportant}
\langle\delta t\rangle = \left({\frac {N}{t_1}+ \frac {N^3}{2t_2}+\frac {N^5}{3t_3}}\right)^{-1},
\end{equation}
plays an important role in the analytical treatment in Sec.~\ref{sub:Analytical}.
\subsection{\label{sub:Classical}Classical master equation approach}

Our stochastic model for the description of the spreading of the center-of-mass density under the influence
of $n$-particle losses ($n=1,2,3$) can be formulated in terms of a classical master equation for the
time-dependent probability distribution $P(X,V,N,t)$, representing the probability density to find
at time $t$ the center of mass coordinate $X$, the corresponding velocity $V$ and the particle
number $N$. Assuming that the various loss events are independent and that
the stochastic process $(X,V,N)$ is Markovian one obtains the
following master equation
\begin{widetext}
\begin{align}
\frac{\partial}{\partial t} P(X,V,N,t) =& -V\frac{\partial}{\partial X} P(X,V,N,t)\hfill 
 + \sum_{n=1}^3 \int dX' \int dV' \Big[ W^{(n)}_{N+n}(X,V|X',V')P(X',V',N+n,t) 
-  W^{(n)}_{N}(X',V'|X,V)P(X,V,N,t) \Big]. 
\label{eq:master}
\end{align}
\end{widetext}
This is a Markovian master equation for a piecewise deterministic process~\cite{Breuer2006}. The first term on the
right-hand side represents the deterministic evolution periods of the center of mass $X$ with
velocity $V$. The deterministic motion is interrupted by random and instantaneous jumps
describing $n$-particles losses, which is described by the second term on the right-hand side.
The transition rate (probability per unit of time) for a jump $X\rightarrow X'$, $V\rightarrow V'$, 
$N\rightarrow N-n$ is
explicitly given by the expression:
\begin{align}
 W^{(n)}_{N}(X',V'|X,V) =& \Gamma^{(n)}_N \sqrt{\frac{1}{2\pi\sigma^2_X(N)}} 
 e^{-\frac{(X-X')^2}{2\sigma^2_X(N)}} \nonumber \\
&\times\sqrt{\frac{1}{2\pi\sigma^2_V(N)}} 
 e^{-\frac{(V-V')^2}{2\sigma^2_V(N)}},
\end{align}
where
\begin{equation}
 \Gamma^{(n)}_N = \frac{N^{2n-1}}{nt_n}, \qquad n=1,2,3.
\end{equation}

As before~\cite{WeissEtAl2015}, $\sigma_V(N)$ and $\sigma_X(N)$ are
related via the uncertainty relation
\begin{equation}
\label{eq:DefSigV}
\sigma_V(N) = \frac {\hbar}{2(N-n)m\sigma_X(N)}.
\end{equation}
While the precise value of  $\sigma^2_X(N)$ remains a fit
parameter for future experiments (or a goal for modeling with a
microscopic model for particle losses), we again choose the rms-width
of a mean-field soliton as the characteristic length-scale~\cite{WeissEtAl2015}
\begin{equation}
\label{eq:DefSigX}
\sigma_X(N)\equiv \frac{\pi\xi_{N-n}}{\sqrt{3}} .
\end{equation}

\section{\label{sec:Results}Results}

\subsection{Overview of Sec.~\ref{sec:Results}}

In section \ref{sub:Analytical} the analytic solution of the model~\cite{WeissEtAl2015} we use to describe the spreading of the center-of-mass density is independent of which type of decoherence via particle losses is implemented. The solution is valid as long as the particle losses are small compared to the total number of particles. Surprisingly, the leading order of the spreading of the center-of-mass density is superballistic, that is the root-mean-square fluctuations of the center-of-mass density scale faster than the \textit{ballistic}\/ prediction
\begin{equation}
\Delta X \propto t;
\end{equation}
the \textit{superballistic}\/ spreading scales as
\begin{equation}
\Delta X \propto t^{3/2}.
\end{equation}
In the following sections we show that the numerics agrees with our analytical prediction and identify parameters for which superballistic motion can be observed experimentally.

\subsection{\label{sub:Analytical}Analytical results, including characteristic time-scales}
In the limit of weak decoherence, the average time per decoherence
event remains roughly constant (rather than increasing with the number
of loss events). Solving the master equation introduced in
Sec.~\ref{sub:Classical} analytically (Appendix~\ref{app:analytically})
yields:
\begin{equation}
\label{eq:XofTpaperleading}
(\Delta X)^2(t) \approx \frac{\sigma_X^2}{\langle\delta
  t\rangle}t + \frac 13\frac{\sigma_V^2}{ \langle\delta t\rangle} t^3,
\end{equation}
Equation~(\ref{eq:XofTpaperleading}) predicts a superballistic spreading of the center-of-mass density of a quantum bright soliton under the influence of decoherence via particle losses --- as long as not too many particles have been lost. In the following subsections, we show that this prediction qualitatively describes the numerics in many parameter regimes: We even find parameters for which the superballistic spreading of the center-of-mass density could be observed in state-of-the art experiments already on short time-scales.

 The point in time where two contributions in
   Eq.~(\ref{eq:XofTpaperleading}) 
 are equal defines a
 characteristic timescale. Together with the definitions at the end of Sec.~\ref{sub:Classical}, it reads
\begin{equation}
\label{eq:tanalyt1}
t^*\equiv\frac{\sqrt{2}\pi^2\hbar^3}{\sqrt{3}m g_{\rm 1D}^2N}.
\end{equation}
Surprisingly, this time-scale is independent of the time-step (strength of decoherence) as long as decoherence is weak --- and is independent of how many particles are lost in one step.

Using Eq.~(\ref{eq:g1d}), Eq.~(\ref{eq:tanalyt1}) can be rewritten to yield
\begin{equation}
\label{eq:tanalyt2}
t^*=\frac{\sqrt{6}\pi^2\hbar}{12m \omega_{\perp}^2Na^2}.
\end{equation}
\textit{For \textsuperscript{7}Li and the experimental parameters of
\cite{KhaykovichEtAl2002}}\footnote{\label{footnote:parametersLi}For \textsuperscript{7}Li, the set of
  parameters used is
  given in Ref.~\cite{KhaykovichEtAl2002} for the \textit{s}-wave scattering
  length~$a= -0.21 \times 10^{-9}\,\rm m$, $\omega_{\perp}=2\pi\times 710\,\rm Hz$. 
For this \textit{s}-wave scattering length we furthermore divide the calculated
  value~\cite{ShotanEtAl2014} for the thermal $K_3$ of $3.6 \times
  10^{-41}\rm m^6/s$ by the factor $3! = 6$ for Bose-Einstein condensates
  and (thus also bright solitons). As we are dealing with ground-state atoms, $K_2=0$ here.}\/ we have
\begin{equation}
\label{eq:timeLi}
t^*_{\rm Li}\simeq\frac{3.4}{N/6000}\, \rm s.
\end{equation}
\textit{For \textsuperscript{85}Rb and the experimental parameters of~\cite{MarchantEtAl2013}}\footnote{\label{footnote:parametersRb}For \textsuperscript{85}Rb is 
  given in Ref.~\cite{MarchantEtAl2013} for the \textit{s}-wave scattering length $a= -11 a_0 = -0.58\times 10^{-9}\,\rm m$, $\omega_{\perp}=2\pi\times 27\,\rm Hz$. For three body-losses, we have $K_3\approx 5 \times 10^{-27} \rm cm^6/s = 5 \times 10^{-39} \rm m^6/s$ and $K_2\approx  3\times 10^{-14}\rm cm^3 / s =  3\times 10^{-20}\rm m^3 / s $~\cite{RobertsEtAl2000}. As described in footnote~\ref{footnote:parametersLi}, for bright solitons we have to divide $K_3$ by $3!$ and additionally have to divide $K_2$ by $2!$.} we find
\begin{equation}
\label{eq:timeRb}
t^*_{\rm Rb}\simeq\frac{25}{N/6000}\, \rm s.
\end{equation}
While we do have 
\begin{equation}
\label{eq:tRb.gt.tLi}
t^*_{\rm Rb} >t^*_{\rm Li}
\end{equation}
for the two parameter-sets given in
footnotes~\ref{footnote:parametersLi} and~\ref{footnote:parametersRb},
there is no principle reason that the time-scale~(\ref{eq:tanalyt2})
has to be larger for Rb bright solitons than for Li bright solitons in
all future experiments. The vertical trapping frequencies are always
likely to be smaller for the heavier Rb-atoms as $m\omega^2x^2$ scales
with the laser intensity used for optical confinement.
 However, $m\omega^2$ is what enters into the equation for the characteristic time~(\ref{eq:tanalyt2}). In the following sections we thus also identify different, experimentally accessible parameter sets for which the characteristic time-scale is considerably shorter.
\begin{figure}[h!]
\includegraphics[width=\linewidth]{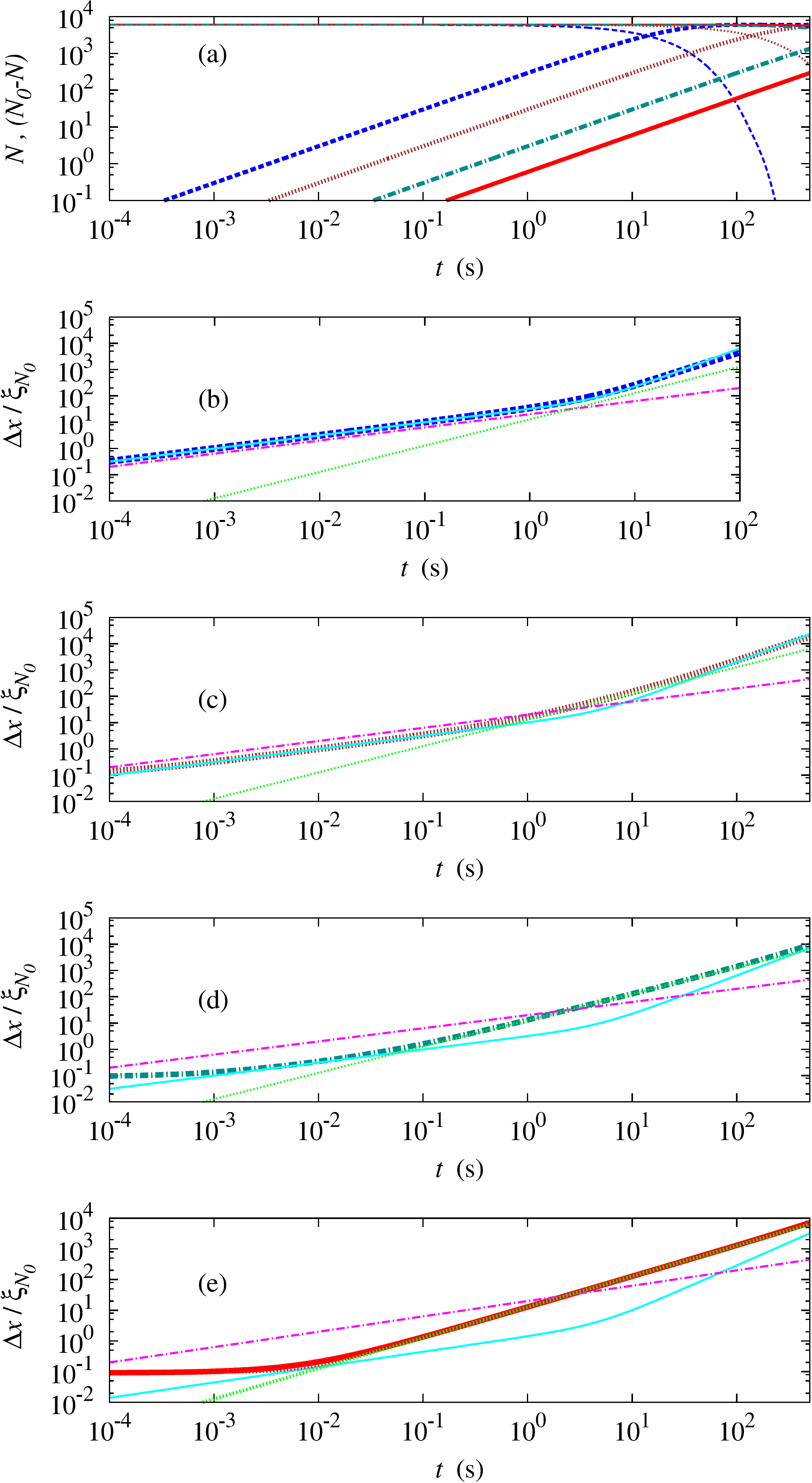}
\caption{\label{fig:single} Li-bright soliton under the influence of single-particle losses (parameters as in footnote~\ref{footnote:parametersLi} but with $K_3=0$ and $N(0)=6000$). Panel a: particle number $N(t)$ (thin curves) and $N(0)-N(t)$ (thick curves). Thick blue (black) dashed curves correspond to a moderate vacuum with single particle losses given by $t_1=20\, \rm s$ (panel b), wide brown (black) short dashed curves an excellent vacuum $t_1=200\, \rm s$ (panel c), wide dark green (black) short dashed curves: $t_1=2000\, \rm s$ (panel d), wide red (black) solid curves: $t_1=10000\, \rm s$ (panel e). Thin light blue (gray) solid curves: analytical formula~(\ref{eq:XofTpaperleading}). As guides to the eye we added the magenta (dark gray) dash-dotted curves ($\propto \sqrt{t}$) and the green (light gray) dotted curves ($\propto t$). Data files are available online~\cite{WeissEtAl2015bData}.} 
\end{figure}

Including the higher-order terms coming from the initial state (cf. Appendix~\ref{app:analytically}),
Eq.~(\ref{eq:XofTpaperleading}) becomes (for not too large times)
\begin{align}
\label{eq:XofTpaper}
(\Delta X)^2(t) \simeq \sigma_{X,0}^2 + \frac{\sigma_X^2}{\langle\delta
  t\rangle}t + \sigma_{V,0}^2t^2 + \frac 13\frac{\sigma_V^2}{ \langle\delta t\rangle} t^3.
\end{align}

\subsection{\label{sec:BrightLi}Bright solitons in  \textsuperscript{7}Li}

As the comparison of the relevant time-scales~(\ref{eq:tRb.gt.tLi}) suggests Li as the more suitable candidate, we start with Li; Rb follows in Sec.~\ref{sec:BrightRb}.

In order to show the validity of the analytical approach we initially
focus on single-particle losses (Fig.~\ref{fig:single}). For the
parameters of the experiment of Ref.~\cite{KhaykovichEtAl2002} (see
footnote~\ref{footnote:parametersLi} but without three-particle
losses), the analytical approach works very well even without the
initial velocity. For the parameters used in Fig.~\ref{fig:single} the
initial velocity only plays an important role for idealized small
values for single particle losses.\footnote{In the appendix in
  Fig~\ref{fig:singleappendix} we show that including the initial
  velocity into the analytical equation considerably increases the
  agreement between our analytical approach and the numerics.}
Superballistic behavior is particularly well visible for less perfect
vacuum. 

\begin{figure}
\includegraphics[width=\linewidth]{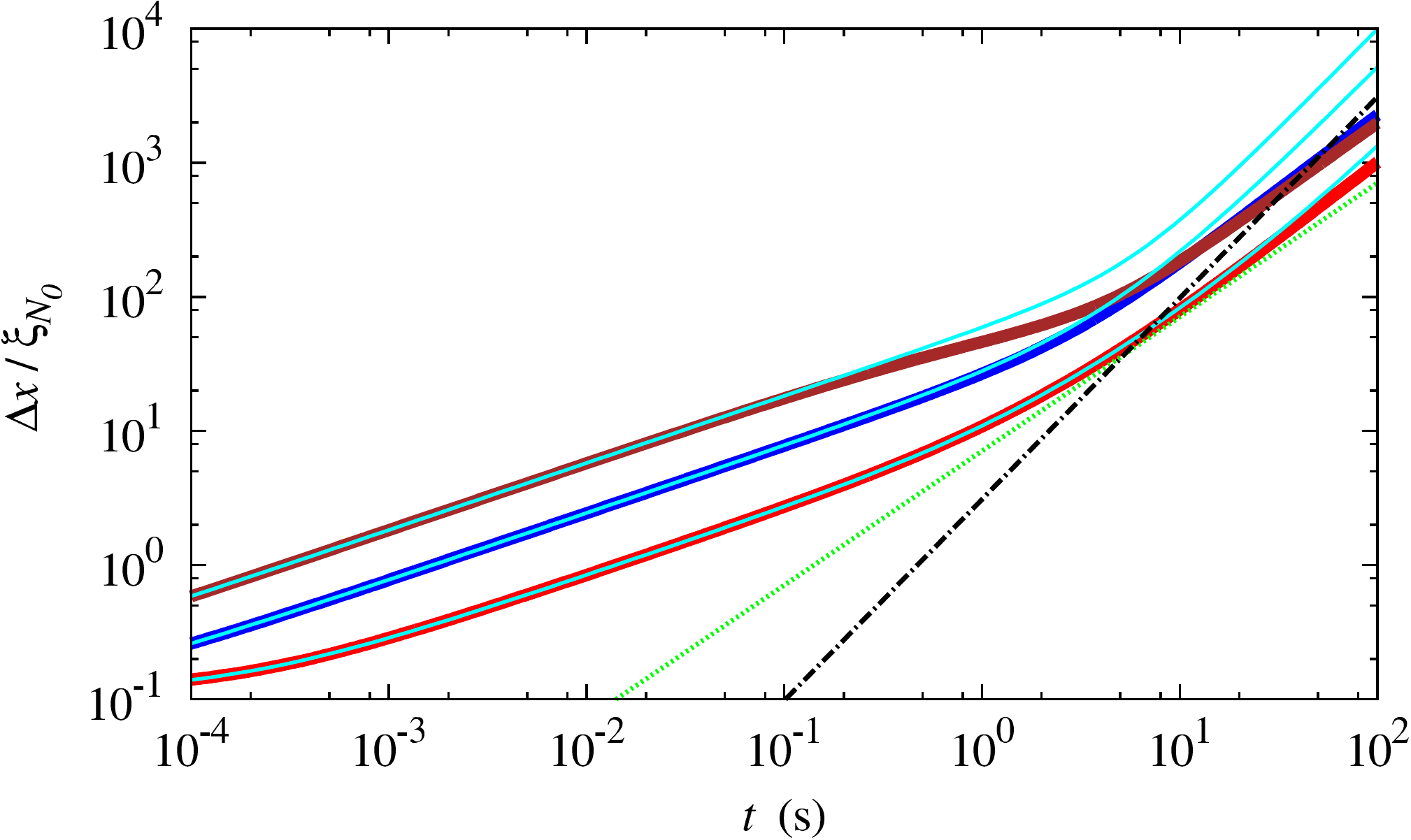}
\caption{\label{fig:old}Root-mean-square fluctuation of the spreading of the center-of-mass density  of a Li-bright soliton as a function of time. Thick
  curves: numerical data from Fig.~3 of Ref.~\cite{WeissEtAl2015}; the
  agreement is good for not too large times. Light blue (dark gray)
  curves:  analytical formula~(\ref{eq:XofTpaper}).  Superballistic
  spreading of the center-of-mass density is barely visible in the numerics and by far not as strong as predicted by the analytical approach. Data files are available online~\cite{WeissEtAl2015bData}.}
\end{figure}

In Fig.~\ref{fig:old} we focus on the dominant three-particle losses as done in Ref.~\cite{WeissEtAl2015}, the initial velocity again only plays a role for some of the parameters. Superballistic spreading of the center-of-mass density is well visible in the analytical curves but only barely visible in the numerics. This clearly indicates that our assumption that the loss rate is constant is not fulfilled. Nevertheless, the analytical equations provide a qualitative understanding for the dynamics.

\begin{figure}
\includegraphics[width=\linewidth]{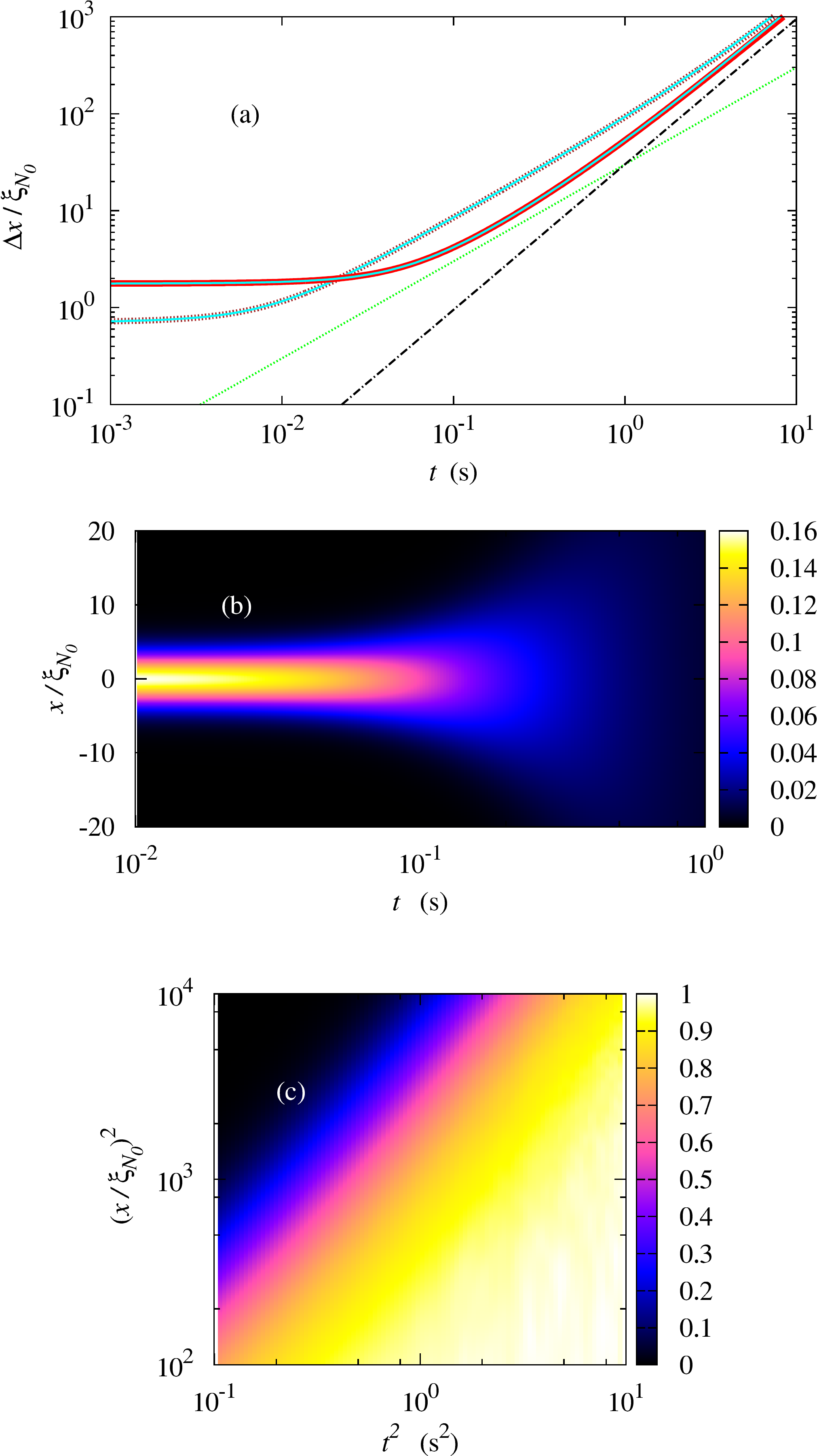}
\caption{\label{fig:super}Superballistic spreading of the center-of-mass density for \textsuperscript{7}Li using
 the parameters of
  footnote~\ref{footnote:parametersLiCWYC} with $N(0)=100$ and a
  moderate vacuum with $t_1=20\, \rm s$. (a)
  Root-mean-square fluctuations of the center-of-mass as a function of
  time. Thick brown (black) dashed
  curve: computer-simulation if the initial harmonic
    oscillator length 
 is 10 soliton lengths. Thick red/black solid curve: weaker initial trap (factor $2.5$ greater harmonic oscillator length) leads to clearly visible superballistic spreading of the center-of-mass density starting earlier. Light blue/light gray dashed and solid curves: corresponding analytical curves~(\ref{eq:XofTpaper}). As guides to the eye we added the green (light gray) dotted curve ($\propto t$) and the black dash-dotted curve ($\propto t^{3/2}$). (b) Two-dimensional projection of the single particle density (which is the convolution of the center-of-mass density and the soliton width) as a function of both time and position. This quantity is experimentally accessible by averaging over the positions of all particles, however it is insightful to plot it differently: by normalizing the maximum to one for each time shown [panel (c)]: Plotting the variance as a function of time squared shows again that the spreading occurs faster than ballistically (which would be parallel to the main diagonal in this panel).
Data files are available online~\cite{WeissEtAl2015bData}.} 
\end{figure}

Unfortunately, superballistic behavior starts rather late. In order to change this, we propose to use the parameters suggested in Ref.~\cite{WeissCastin2009}.\footnote{\label{footnote:parametersLiCWYC}For \textsuperscript{7}Li and $N\approx 100$, the set of
  parameters used is
  given in Ref.~\cite{WeissCastin2009} for the \textit{s}-wave scattering
  length~$a= -1.72 \times 10^{-9}\,\rm m$, $\omega_{\perp}=2\pi\times 4800\,\rm Hz$. 
  For $K_3$ and $K_2$ we use the parameters given in footnote~\ref{footnote:parametersLi}; for practical purposes and the moderate vacuum used in Fig~\ref{fig:super} we could have set $K_3=0$ (in addition to setting $K_2=0$).
} If the value of the initial trap has a harmonic
  oscillator length $\sqrt{\hbar/(m\omega)}$
that is 10 larger than the soliton length $\xi_N$ (the value used in
all other figures), Fig.~\ref{fig:super}, primarily shows ballistic
spreading of the center-of-mass density. However, as predicted by the analytical
approach~(\ref{eq:XofTpaper}), using an initial trap for which the
harmonic oscillator length is $25$ soliton lengths,
 superballistic spreading of the center-of-mass density becomes clearly visible already at short time-scales.

\subsection{\label{sec:BrightRb}Bright solitons in  \textsuperscript{85}Rb}

\begin{figure}
\includegraphics[width=\linewidth]{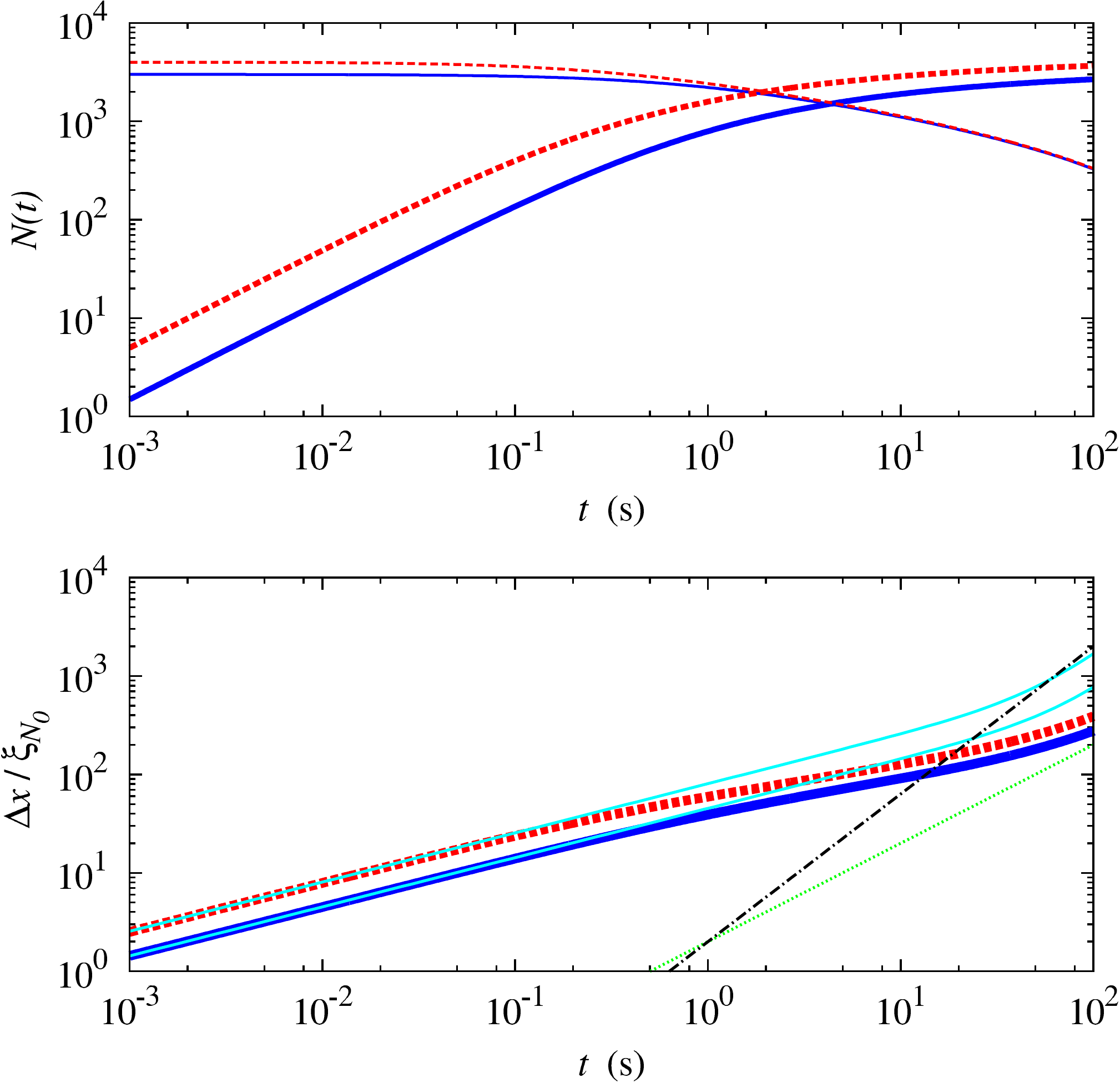}
\caption{\label{fig:Rb} Rb-bright solitons under the influence of
  three-, two- and single-particle losses. Most parameters used can be
  found in footnote~\ref{footnote:parametersRb}, additionally a very
  good vacuum with $t_1= 200\, \rm s$ was chosen. \textit{Upper
    panel}: red (black) dashed curves correspond to $N_0=4000$, blue
  (black) solid curves correspond to $N_0= 3000$.  Thin curves:
  $N(t)$; thick curves: $N_0-N(t)$. \textit{Lower panel}:
  Root-mean-square fluctuations of the center-of-mass position.  Red
  (black) dashed curves correspond to $N_0=4000$, blue (black) solid
  curves correspond to $N_0= 3000$. Light blue (light gray) solid
  curves: analytical curves that assume the time $\langle \delta
  t\rangle$ for one loss event remains constant [Eq.~(\ref{eq:XofTpaper})]. As guides to the eye for ballistic motion a curve $\propto t$ [green (light gray) dotted line] and  a curve   $\propto t^{3/2}$ for superballistic motion (black dash-dotted line) have been added. Data files are available online~\cite{WeissEtAl2015bData}.}
\end{figure}

Let us start by comparing the time-scales for Li and Rb bright solitons using the parameters in footnotes~\ref{footnote:parametersLi} and \ref{footnote:parametersRb}, based on the experiments of Refs.~\cite{KhaykovichEtAl2002} and \cite{MarchantEtAl2013}, Eqs.~(\ref{eq:timeLi}) and (\ref{eq:timeRb}). Figure~\ref{fig:Rb} confirms that Rb-bright solitons are less useful to investigate superballistic spreading of the center-of-mass density than Li-bright solitons if one uses the experimental parameters of Refs.~\cite{KhaykovichEtAl2002} and \cite{MarchantEtAl2013}: even if we chose an excellent vacuum, superballistic spreading of the center-of-mass density is not observable as too many particles are lost already.

\begin{figure}
\includegraphics[width=\linewidth]{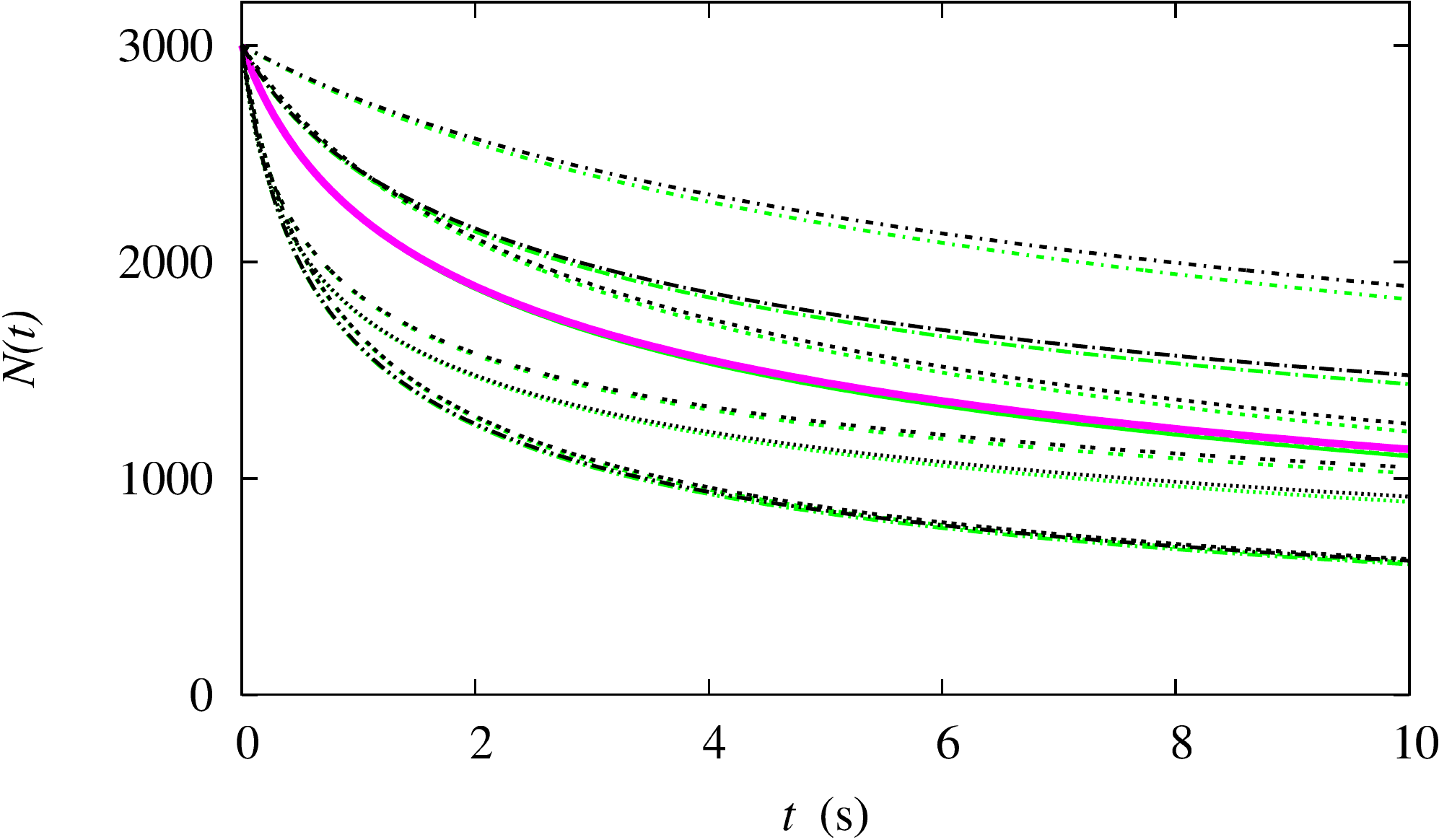}
\caption{\label{fig:RbNt} Number of atoms in a Rb-bright solitons under the influence of three-, two- and single-particle losses for  $N=3000$  and the parameters given in footnote~\ref{footnote:parametersRb}. Thick magenta (dark gray) line: parameters as in footnote~\ref{footnote:parametersRb}. Thin black lines: an error of a factor of 5 was added to the loss parameters. Green (light gray) lines: also includes single-particle losses with  $t_1= 200\, \rm s$. All data generated with analytical equations~\cite{maple} [cf.~Eqs.~(\ref{eq:dt}) and (\ref{eq:Nana})]; the numerical data from Fig.~\ref{fig:Rb} lies on top of the corresponding green curve in this figure (which in turn is partially identical to the thick magenta (dark gray) curve). Data files are available online~\cite{WeissEtAl2015bData}.}
\end{figure}

However, even without changing the experimental parameters in future
Rb-experiments as suggested in the lines below
Eq.~(\ref{eq:tRb.gt.tLi}), performing such experiments can be very
useful. Contrary to the case of Li, both two-particle and
three-particle losses are present for Rb. If we assume that the values
given in footnote~\ref{footnote:parametersRb} have an error of a
factor of 5, this leads to quite distinct curves for the number of
atoms as a function of time (Fig.~\ref{fig:RbNt}). Contrary to the
experiment of Ref.~\cite{RobertsEtAl2000} for which two-particle
losses are the dominant loss process, for the bright solitons
investigated experimentally in~\cite{MarchantEtAl2013} both loss rates
are initially comparable. The effects of single particle
  losses would have to be included only for a very much smaller error
  margin.

If, on the other hand, we go the path of changing the parameters in the Rb-experiments~\cite{MarchantEtAl2013,MarchantEtAl2016}, one approach would be to choose deep optical lattices perpendicular to the quasi-one-dimensional wave guide which would allow trapping frequencies in the kHz regime. Implementing optical lattices might even provide the possibility of having many tubes in which a very similar experiment is performed, thus allowing to average over different realizations of the spreading of the center-of-mass density in a single experiment.

\begin{figure}
\includegraphics[width=\linewidth]{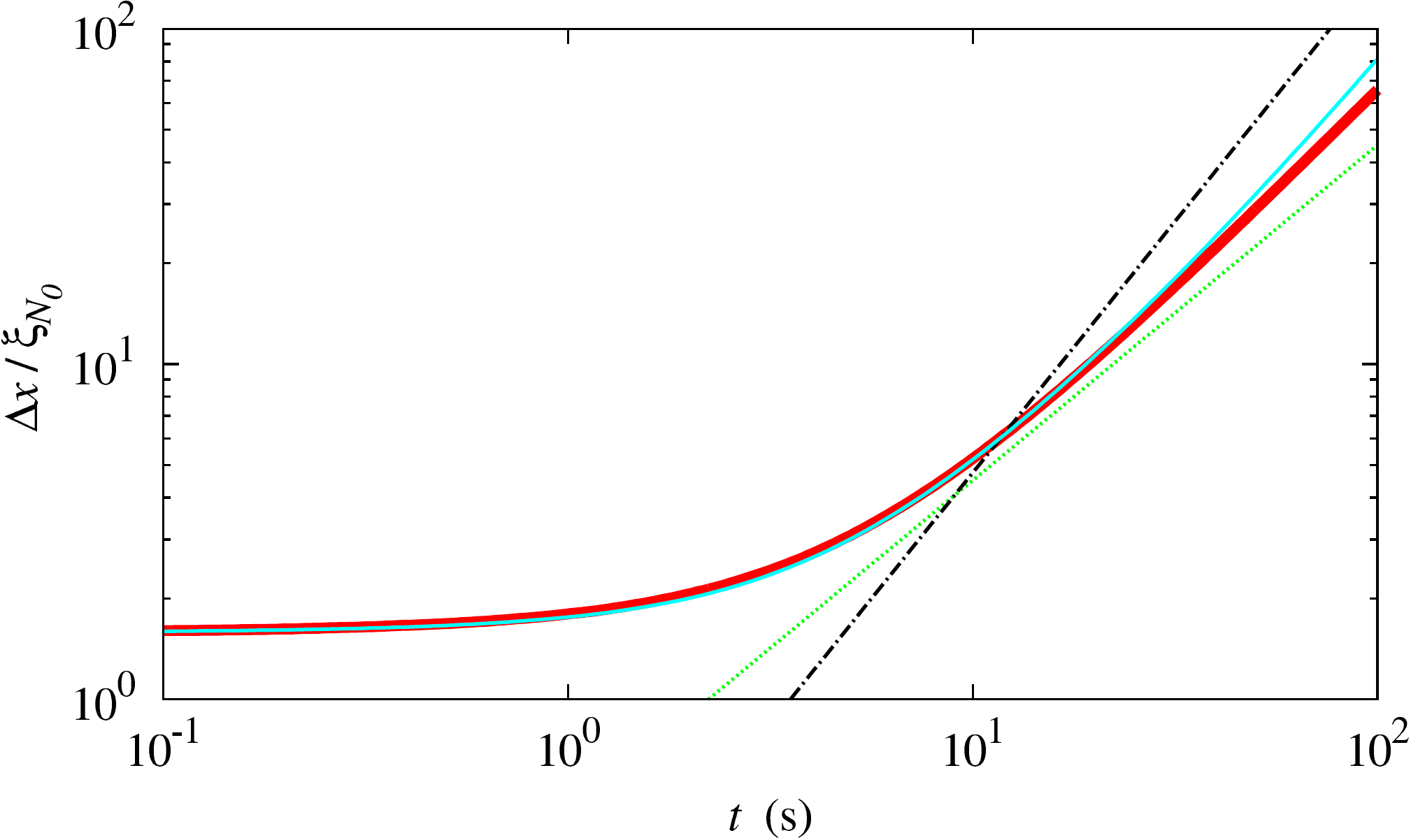}
\caption{\label{fig:superRb}Superballistic spreading of the center-of-mass density in \textsuperscript{85}Rb-bright solitons. The parameters are the same as in footnote~\ref{footnote:parametersRb} except for $\omega_{\perp} = 2 \pi \times 0.972\,\rm kHz$, the particle number is lower than in all other plots ($N=20$) and the vacuum is nearly perfect ($t_1 = 2000\,\rm s$). As for nearly all other curves, the initial trap is a factor of 10 larger than the soliton length. While the numerics [thick red (black) curve] does not reach the $\propto t^{3/2}$ behavior (black dash-dotted curve), predicted by the analytical approach [light blue (light gray) solid line; Eq.~(\ref{eq:XofTpaper})] it does grow faster than $\propto t$ [green (light gray) dotted line]. Data files are available online~\cite{WeissEtAl2015bData}.}
\end{figure}

For Fig.~\ref{fig:superRb}, we use the parameters of footnote~\ref{footnote:parametersRb} except for $\omega_{\perp} = 2 \pi \times 0.972\,\rm kHz$. This increase of the trapping frequency by a factor of 36 reduces the perpendicular harmonic oscillator length only by a factor of 6 while reducing the soliton length (\ref{eq:solitonlength}) via Eq.~(\ref{eq:g1d}) by a factor of 36 (if $N$ remains of the order of 6000 atoms). While this endangers the one-dimensional character of our wave guide, this can easily be compensated by reducing the particle numbers. We thus reduce the particle number. When doing this, we also have to ensure that $10\times t^*/\langle \delta t\rangle \ll N_0$ is fulfilled (to be in the regime of weak decoherence even after superballistic spreading of the center-of-mass density  has set in, thus we have to fulfill [cf.~Eqs.~(\ref{eq:t3}), (\ref{eq:t2}), (\ref{eq:tanalyt1}) and (\ref{eq:timportant})]
\begin{equation}
\label{eq:superballisticrequirement}
10t^* \left({\frac {1}{t_1}+ \frac {N^2}{2t_2}+\frac {N^4}{3t_3}}\right) \ll 1
\end{equation}
The fact that three-body losses are larger for Rb than for Li (see footnotes~\ref{footnote:parametersLi} and \ref{footnote:parametersRb}) requires low particle numbers to make the second and third term small, as $t^*\propto \frac 1 N$, the first term then requires nearly perfect vacuum. As a proof of principle, Fig.~\ref{fig:superRb} displays superballistic spreading of the center-of-mass density for a Rb-bright soliton. However, contrary to what we suspected in Ref.~\cite{WeissEtAl2015} it is not the heavier mass of Rb that makes it less useful for experimental realizations --- the analytic treatment leading to Eq.~(\ref{eq:superballisticrequirement}) shows that it is rather the higher loss rates. While the time-scale in Fig.~\ref{fig:superRb} could easily be reduced by choosing higher particle numbers, two- and three- particle losses would then prevent us from observing superballistic spreading of the center-of-mass density in both computer simulations and experiments.

\section{\label{sec:Conclusion}Conclusion and outlook}

To conclude, the main results of our paper treating attractively interacting Bosons in a quasi-one-dimensional waveguide with an additional initial weak harmonic trap  are:
\begin{enumerate}
\item We present an analytical solution
for the numerical model of the spreading of the center-of-mass density introduced in Ref.~\cite{WeissEtAl2015} under the influence of decoherence via single- two- or three-body losses.
\item For stronger decoherence, the analytical
model still qualitatively describes the transition from short-time
diffusive to long-time ballistic behavior investigated numerically in
Ref.~\cite{WeissEtAl2015} (Figs.~\ref{fig:old} and \ref{fig:Rb}).
\item The
analytical solutions predict center-of-mass rms fluctuations as a function
of time that scales as $\propto t^{3/2}$; in the numerics the scaling is slower
but still considerably faster than the ballistic ($\propto t$) regime (Figs.~\ref{fig:super} and
\ref{fig:superRb}).
\item  For \textsuperscript{85}Rb, measuring
the decay of the number
of particles could furthermore help narrowing down the error margins
for two- and three-particle losses (Fig.~\ref{fig:RbNt}).
\end{enumerate}

 For many
aspects of the spreading of the center-of-mass density \textsuperscript{7}Li-bright
solitons are more suitable as, in particular, the
 time-scale for particle
losses is longer. Our model differs considerably from the noise-driven
motion of Ref.~\cite{MaucherEtAl2012} and other systems used to
investigate superballistic motion (see~\cite{ZhaoEtAl2014} and
references therein): The decoherence-induced spreading of the center-of-mass density of quantum bright
solitons described via the numerical model of
Ref.~\cite{WeissEtAl2015} can be viewed as a mesoscopic signature of
microscopic quantum physics. The analytic solution presented here
allowed us to predict and subsequently numerically observe
superballistic motion.

Decoherence via particle losses is also likely to affect predictions beyond the center-of-mass motion. Unless one uses the approach of Ref.~\cite{WeissCastin2009} to focus on experiments with time-scales shorter than the first decoherence-event, theoretical predictions for bright solitons are likely to change if decoherence via particle losses is included.  

Topics for which this might play a role include interferometric
applications~\cite{PoloAhufinger2013,StreltsovaStreltsov2014,HelmEtAl2015}
and modeling the  collisions of two bright solitons observed recently
in the experiment of Ref.~\cite{NguyenEtAl2014}
(cf.~\cite{Martin2016}) --- in particular as soon as beyond-mean field
quantum effects play a role~\cite{HoldawayEtAl2013} in these
collisions. The long-time behavior of bright solitons after scattering
from a barrier, investigated experimentally for a large repulsive
barrier in Ref.~\cite{MarchantEtAl2013} and for a narrow attractive
barrier in Ref.~\cite{MarchantEtAl2016}, are likely to be affected.\footnote{The barriers used, for example, in Refs.~\cite{MarchantEtAl2013,MarchantEtAl2016} were made with a laser focus. For more complex structures written with light that could be used for experiments with ultra-cold atoms see Ref.~\cite{BowmanEtAl2015}.}

The model introduced in Ref.~\cite{WeissEtAl2015} and solved analytically in the current paper is based on the unique properties of quantum bright solitons. Developing a similar model valid for repulsive interactions is an interesting question for future research.

The data presented in this paper 
will be available
 online at~\url{http://dx.doi.org/10.15128/44558d350}~\cite{WeissEtAl2015bData}.

\acknowledgments

We thank
 S.\ A.\ Hopkins, V.\ M.\ Kendon and  L.\ Khaykovich for
discussions. C.W.\ thanks the \textit{Physikalisches Institut}, \textit{Universit\"at Freiburg}, Germany, for its hospitality.  We thank the UK Engineering and Physical Sciences Research Council (Grant No.\ EP/L010844/1, C.W., S.L.C and
S.A.G.) for funding.

\begin{appendix}

\section{\label{app:LiebLiniger}Lieb-Linger model with attractive interactions}
For attractively interacting atoms ($g_{\rm 1D}<0$) in one dimension, the Lieb-Liniger-(McGuire) Hamiltonian~\cite{LiebLiniger1963,McGuire1964} is a very useful model
\begin{equation}
\hat{H} = -\sum_{j=1}^N\frac{\hbar^2}{2m}\frac{\partial^2}{\partial x_j^2}+\sum_{j=1}^{N-1}\sum_{n=j+1}^{N}g_{\rm 1D}\delta(x_j-x_n);
\label{eq:LL}
\end{equation}
 $x_j$ denotes the position of particle $j$ of mass $m$. For this model, even the (internal) ground state wave function is known analytically. Including the center-of-mass momentum $K$, the corresponding eigenfunctions relevant for our dynamics read (cf.~\cite{CastinHerzog2001})
\begin{equation}
\Psi(x_1,x_2,\ldots,x_N) \propto e^{iKX} \exp\left(-\frac{m |g_{\rm 1D}|}{2\hbar^2}\sum_{j<\nu}|x_j-x_{\nu}|\right);
\end{equation}
the center-of-mass coordinate is given by Eq.~(\ref{eq:CoMCoord}). If the center-of-mass wave function is a delta function and the particle number is $N\gg 1$, then the single-particle density can be shown~\cite{CalogeroDegasperis1975,CastinHerzog2001} to be equivalent to the mean-field result~(\ref{eq:singlesoliton}). Thus, the Lieb-Liniger model is a one-dimensional many-particle quantum model that can be used to justify the approach to treat a quantum bright soliton like a mean-field soliton with additional center-of-mass motion after opening a weak initial trap. In the limit $N\to\infty$, $g_{\rm 1D}\to 0$ such that $Ng_{\rm 1D}= \rm const$, the initial width of the center-of-mass wave function goes to zero, $\Delta X_0\propto 1/\sqrt{N}$.

\section{Deriving the analytic results\label{app:analytically}}

In order to derive an analytical expression for the variance of the position of the center
of mass $X$ we use the approximation of a constant particle number $N$. The master
equation (\ref{eq:master}) can then be written in the simpler form
\begin{widetext}
\begin{align} \label{MEQ}
 \frac{\partial}{\partial t} P(X,V,t) =& -V\frac{\partial}{\partial X} P(X,V,t)
 + \int dX' \int dV' W(X-X',V-V')P(X',V',t) - \Gamma P(X,V,t), 
\end{align}
where $P(X,V,t)$ is the probability to find at time $t$ the center of mass coordinate $X$
and the velocity $V$. The rate for a transition  $X \rightarrow X'$, $V \rightarrow V'$ is given by
\begin{equation}
 W(X-X',V-V') = \Gamma \sqrt{\frac{1}{2\pi\sigma^2_X}} 
 \exp\left({-\frac{(X-X')^2}{2\sigma^2_X}}\right) \sqrt{\frac{1}{2\pi\sigma^2_V}} 
 \exp\left({-\frac{(V-V')^2}{2\sigma^2_V}}\right),
\end{equation}
\end{widetext}
and the total transition rate takes the form
\begin{equation}
 \Gamma = \Gamma^{(1)}+\Gamma^{{2}}+\Gamma^{(3)} 
 = \frac{N}{t_1} + \frac{N^3}{2t_2} + \frac{N^5}{3t_3} 
 = \langle \delta t \rangle^{-1}.
\end{equation}
From the master equation (\ref{MEQ}) one can derive, \textit{without further approximations}, 
the following equations of motion for the first and second moments of the process: 
\begin{eqnarray}
 \frac{d}{dt} \langle X(t) \rangle &=& \langle V(t) \rangle, \label{EQ-MEAN-X} \\
 \frac{d}{dt} \langle V(t) \rangle &=& 0, \label{EQ-MEAN-V} \\ 
 \frac{d}{dt} \langle X^2(t) \rangle &=& 2\langle X(t)V(t) \rangle + \Gamma\sigma^2_X, \\
 \frac{d}{dt} \langle V^2(t) \rangle &=& \Gamma\sigma^2_V, \\
 \frac{d}{dt} \langle X(t)V(t) \rangle &=& \langle V^2(t) \rangle. \label{EQ-MEAN-XV}
\end{eqnarray}
For example, to derive Eq. (\ref{EQ-MEAN-X}) one starts from
\begin{equation}
 \langle X(t) \rangle = \int dX \int dV X P(X,V,t),
\end{equation}
and takes the time derivative:
\begin{equation}
 \frac{d}{dt} \langle X(t) \rangle = \int dX \int dV X \frac{\partial}{\partial t} P(X,V,t).
\end{equation}
Substituting the master equation (\ref{MEQ}) leads to:
\begin{widetext}
\begin{align} 
 \frac{d}{dt} \langle X(t) \rangle =- \int dX \int dV XV\frac{\partial}{\partial X} P(X,V,t) 
 + \int dX \int dV \int dX' \int dV' X W(X-X',V-V') P(X',V',t) 
  - \Gamma \langle X(t) \rangle. 
\end{align}
\end{widetext}
After partial integration the first term on the right-hand side yields $\langle V(t) \rangle$.
Integrating first over $X$ and $V$ the second term gives $+\Gamma \langle X(t) \rangle$
which cancels out the third term. This leads to Eq. (\ref{EQ-MEAN-X}).
In a similar way Eqs. (\ref{EQ-MEAN-V}) - (\ref{EQ-MEAN-XV}) can be obtained.

The closed system of differential equations (\ref{EQ-MEAN-X}) - (\ref{EQ-MEAN-XV})
for the moments can easily be solved to yield:
\begin{align} \label{Variance-X}
 \langle X^2(t) \rangle - \langle X(t) \rangle^2 =&
 \sigma^2_{X,0} + \Gamma \sigma_X^2 \; t + \sigma^2_{V,0} \; t^2 
 + \frac{1}{3} \Gamma \sigma_V^2 \; t^3 \nonumber \\ 
 & + 2\Big(\langle X(0)V(0)\rangle - \langle X(0)\rangle \langle V(0) \rangle\Big) \, t,
\end{align}
where
\begin{align}
 \sigma^2_{X,0} &= \langle X^2(0) \rangle - \langle X(0)\rangle^2, \\
 \sigma^2_{V,0} &= \langle V^2(0) \rangle - \langle V(0)\rangle^2. 
\end{align}
The last term on the right-hand side of Eq. (\ref{Variance-X}) does not appear in the main
text as it is zero because position and velocity are uncorrelated at
the initial time.\footnote{The full quantum mechanical expression for the last term on the right-hand side of Eq. (\ref{Variance-X})
  reads: $2[\left(\langle X(0)V(0)\rangle +\langle X(0)V(0)\rangle\right)/2 - \langle X(0)\rangle \langle V(0)\rangle$]t. }

\section{Random walk in velocity space\label{app:velocity}}

For a random-walk in velocity space~\cite{Obukhov1959,BauleFriedrich2006,KesslerBarkai2012} with Gaussian step-distribution characterized by
\begin{align}
\label{eq:sigmaV}
  \sigma_V &= \frac{\hbar}{2(N-3)m\sigma_X}
\end{align}
where 
\begin{align}
\sigma_X &= \frac{\pi \xi_N}{\sqrt{3}}= \frac{\pi \hbar^2}{\sqrt{3}m|g_{\rm 1D}|(N-4)}
\end{align}
leads, for $(N-4)/(N-3)\simeq 1$, to an $N$- and particle-mass independent step-size:
\begin{equation}
\sigma_V= \frac{\sqrt{3}|g_{\rm 1D}|}{2\pi\hbar}\;.
\end{equation}
For the velocity after $n$ random-walk steps we thus have:
\begin{equation}
(V)_n= \sum_{\ell=1}^n\delta V_{\ell}
\end{equation}
For an $n$-independent time-step $\delta t$ (thus assuming $N\simeq N-n$), we have:
\begin{align}
\label{eq:approxV0}
(\Delta X)^2_{n~V~\rm steps}  &\equiv \langle X(t)^2\rangle_{n~V~\rm steps} \\\nonumber
&=\rangle\left<\left(\sum_{\nu=1}^{n}V_{\nu}\right) \left(\sum_{\mu=1}^{n}V_{\mu}\right) \langle \delta t_{\mu}\delta t_{\nu}\rangle\right>_{n~V~\rm steps}\\\nonumber
&=\langle\delta t \rangle^2\sum_{\nu=1}^{n}\sum_{\mu=1}^{n}\sum_{\ell=1}^{\nu}\sum_{j=1}^{\mu}\left<\delta V_{\ell}\delta V_{j}\right>\\\nonumber
&=\langle \delta t \rangle^2\sum_{\nu=1}^{n}\sum_{\ell=1}^{\nu}\sum_{\mu=1}^{n}\sum_{j=1}^{\mu}\sigma_V^2\delta_{\ell,j}\\\nonumber
&=\langle \delta t \rangle^2\sigma_V^2\sum_{\nu=1}^{n}\sum_{\ell=1}^{\nu}\sum_{\mu=\ell}^{n}1.
\end{align}
Solving the remaining sums analytically yields~\cite{maple}
\begin{align}
\label{eq:rmsXvspace}
(\Delta X)^2_{n~V~\rm steps}  & = \left(\frac 13 {n}^{3}+\frac 12{n}^{2}+\frac 16n\right)\langle \delta t \rangle^2\sigma_V^2\\
\label{eq:leading}
& \simeq \frac 13 {n}^{3}\langle \delta t \rangle^2\sigma_V^2.
\end{align}

The above assumes that $\langle\delta t^2\rangle$ is $n$-independent;
the $(\Delta X)^2\propto t^3$ dependence is visible because of $n
\propto t$.

\section{Estimating the initial velocity}
\begin{figure}
\includegraphics[width=\linewidth]{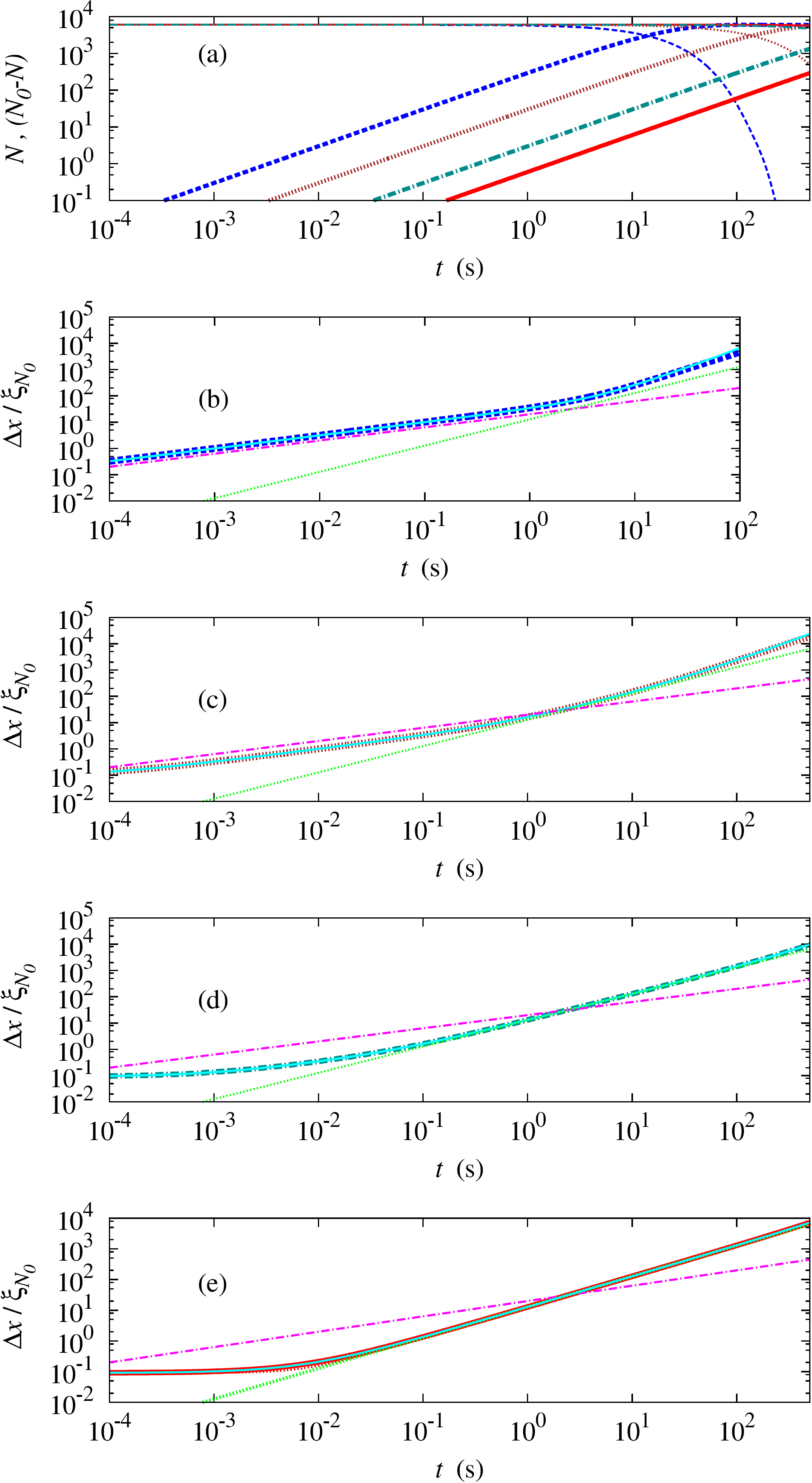}
\caption{\label{fig:singleappendix} Li-bright soliton under the
  influence of single-particle losses (parameters as in
  footnote~\ref{footnote:parametersLi} but with $K_3=0$ and
  $N(0)=6000$). The data are the same as for Fig.~\ref{fig:single} but
  for the fact that the analytical curves now include the initial
  velocity and the initial width of the wavepacket [Eq.~(\ref{eq:XofTpaper})]. This leads to a much better agreement for weaker decoherence. Data files are available online~\cite{WeissEtAl2015bData}.}
\end{figure}

Figure~\ref{fig:singleappendix} shows the importance of including the initial velocity: If the initial velocity is added to the analytical curves depicted in Fig.~\ref{fig:single}, this considerably increases the agreement between analytical and numerical results.
Comparing the very good agreement between analytical and numerical
results if the correct value of the initial velocity is used
(Fig.~\ref{fig:singleappendix}) to the approximation $\sigma_{V,0}=0$
(and $\sigma_{X,0}=0$) of Fig.~\ref{fig:single} shows that the initial
velocity does indeed play a role and merits our attention.

Particle losses are particularly easy to model if we have a product
state. We start with a noninteracting Bose gas in the ground state of a one-dimensional harmonic trap; both in position space and in velocity space we have:
\begin{equation}
\label{eq:sigapp}
\sigma^2_N = \frac{\sigma^2_1}N
\end{equation}
and this changes to 
\begin{equation}
\sigma^2_{N-\nu} = \frac{\sigma^2_1}{N-\nu}
\end{equation}
after one loss event losing $\nu$ particles, thus increasing the variance by
\begin{equation}
\label{eq:onestep}
\widetilde{\sigma}^2 =  \sigma^2_1\left(\frac{1}{N-\nu}-\frac{1}{N}\right).
\end{equation}
In order to estimate how long our assumption that the
  initial velocity distribution is given by Eq.~(\ref{eq:sigapp})
  remains valid, we use a linear variation of the additional variance introduced in one step~(\ref{eq:onestep}) during the ramping process
\begin{equation}
 \sigma^2(t) =  \widetilde{\sigma}^2 +  (\sigma^2_{\rm our~model}-\widetilde{\sigma}^2)\frac tT.
\end{equation}
For a specific experiment, we thus can check if 
\begin{equation}
\frac 1{\langle \delta t\rangle}\int_{0}^T \sigma^2(t) \ll \frac{\sigma^2_1}N
\end{equation}
is indeed fulfilled. With $T$ typically in the tens of milliseconds~\cite{Khaykovich2015} for experiments like~\cite{KhaykovichEtAl2002}, $T/\langle\delta t\rangle \approx 50\, {\rm ms} / (200\, {\rm s}/N)$ if single-particle losses are the dominant source of decoherence during the adiabatic switching. For $N=6000$ we have less than 2 loss events and thus do not have to change the initial velocity in our model. The larger trapping frequencies for Li as compared to the heavier Rb leads to shorter switching times for Li. While this again is an argument for choosing lighter atoms for this type of experiment, future experiments are likely to show if further modeling is necessary.

\end{appendix}

%


\begin{thebibliography}{85}%
\makeatletter
\providecommand \@ifxundefined [1]{%
 \@ifx{#1\undefined}
}%
\providecommand \@ifnum [1]{%
 \ifnum #1\expandafter \@firstoftwo
 \else \expandafter \@secondoftwo
 \fi
}%
\providecommand \@ifx [1]{%
 \ifx #1\expandafter \@firstoftwo
 \else \expandafter \@secondoftwo
 \fi
}%
\providecommand \natexlab [1]{#1}%
\providecommand \enquote  [1]{``#1''}%
\providecommand \bibnamefont  [1]{#1}%
\providecommand \bibfnamefont [1]{#1}%
\providecommand \citenamefont [1]{#1}%
\providecommand \href@noop [0]{\@secondoftwo}%
\providecommand \href [0]{\begingroup \@sanitize@url \@href}%
\providecommand \@href[1]{\@@startlink{#1}\@@href}%
\providecommand \@@href[1]{\endgroup#1\@@endlink}%
\providecommand \@sanitize@url [0]{\catcode `\\12\catcode `\$12\catcode
  `\&12\catcode `\#12\catcode `\^12\catcode `\_12\catcode `\%12\relax}%
\providecommand \@@startlink[1]{}%
\providecommand \@@endlink[0]{}%
\providecommand \url  [0]{\begingroup\@sanitize@url \@url }%
\providecommand \@url [1]{\endgroup\@href {#1}{\urlprefix }}%
\providecommand \urlprefix  [0]{URL }%
\providecommand \Eprint [0]{\href }%
\providecommand \doibase [0]{http://dx.doi.org/}%
\providecommand \selectlanguage [0]{\@gobble}%
\providecommand \bibinfo  [0]{\@secondoftwo}%
\providecommand \bibfield  [0]{\@secondoftwo}%
\providecommand \translation [1]{[#1]}%
\providecommand \BibitemOpen [0]{}%
\providecommand \bibitemStop [0]{}%
\providecommand \bibitemNoStop [0]{.\EOS\space}%
\providecommand \EOS [0]{\spacefactor3000\relax}%
\providecommand \BibitemShut  [1]{\csname bibitem#1\endcsname}%
\let\auto@bib@innerbib\@empty
\bibitem [{\citenamefont {Zaburdaev}\ \emph {et~al.}(2008)\citenamefont
  {Zaburdaev}, \citenamefont {Schmiedeberg},\ and\ \citenamefont
  {Stark}}]{ZaburdaevEtAl2008}%
  \BibitemOpen
  \bibfield  {author} {\bibinfo {author} {\bibfnamefont {V.}~\bibnamefont
  {Zaburdaev}}, \bibinfo {author} {\bibfnamefont {M.}~\bibnamefont
  {Schmiedeberg}}, \ and\ \bibinfo {author} {\bibfnamefont {H.}~\bibnamefont
  {Stark}},\ }\bibfield  {title} {\enquote {\bibinfo {title} {Random walks with
  random velocities},}\ }\href {\doibase 10.1103/PhysRevE.78.011119} {\bibfield
   {journal} {\bibinfo  {journal} {Phys. Rev. E}\ }\textbf {\bibinfo {volume}
  {78}},\ \bibinfo {pages} {011119} (\bibinfo {year} {2008})}\BibitemShut
  {NoStop}%
\bibitem [{\citenamefont {Zimbardo}\ \emph {et~al.}(2000)\citenamefont
  {Zimbardo}, \citenamefont {Greco},\ and\ \citenamefont
  {Veltri}}]{ZimbardoEtAl2000}%
  \BibitemOpen
  \bibfield  {author} {\bibinfo {author} {\bibfnamefont {G.}~\bibnamefont
  {Zimbardo}}, \bibinfo {author} {\bibfnamefont {A.}~\bibnamefont {Greco}}, \
  and\ \bibinfo {author} {\bibfnamefont {P.}~\bibnamefont {Veltri}},\
  }\bibfield  {title} {\enquote {\bibinfo {title} {Superballistic transport in
  tearing driven magnetic turbulence},}\ }\href {\doibase
  http://dx.doi.org/10.1063/1.873914} {\bibfield  {journal} {\bibinfo
  {journal} {Physics of Plasmas}\ }\textbf {\bibinfo {volume} {7}},\ \bibinfo
  {pages} {1071} (\bibinfo {year} {2000})}\BibitemShut {NoStop}%
\bibitem [{\citenamefont {Prants}\ \emph {et~al.}(2002)\citenamefont {Prants},
  \citenamefont {Edelman},\ and\ \citenamefont {Zaslavsky}}]{PrantsEtAl2002}%
  \BibitemOpen
  \bibfield  {author} {\bibinfo {author} {\bibfnamefont {S.~V.}\ \bibnamefont
  {Prants}}, \bibinfo {author} {\bibfnamefont {M.}~\bibnamefont {Edelman}}, \
  and\ \bibinfo {author} {\bibfnamefont {G.~M.}\ \bibnamefont {Zaslavsky}},\
  }\bibfield  {title} {\enquote {\bibinfo {title} {Chaos and flights in the
  atom-photon interaction in cavity {QED}},}\ }\href {\doibase
  10.1103/PhysRevE.66.046222} {\bibfield  {journal} {\bibinfo  {journal} {Phys.
  Rev. E}\ }\textbf {\bibinfo {volume} {66}},\ \bibinfo {pages} {046222}
  (\bibinfo {year} {2002})}\BibitemShut {NoStop}%
\bibitem [{\citenamefont {Liu}\ and\ \citenamefont {Bao}(2014)}]{LiuBao2014}%
  \BibitemOpen
  \bibfield  {author} {\bibinfo {author} {\bibfnamefont {Y.}~\bibnamefont
  {Liu}}\ and\ \bibinfo {author} {\bibfnamefont {J.~D.}\ \bibnamefont {Bao}},\
  }\bibfield  {title} {\enquote {\bibinfo {title} {Generation and application
  of non-ergodic noise},}\ }\href {\doibase 10.7498/aps.63.240503} {\bibfield
  {journal} {\bibinfo  {journal} {Acta Physica Sinica}\ }\textbf {\bibinfo
  {volume} {63}},\ \bibinfo {eid} {240503} (\bibinfo {year}
  {2014})}\BibitemShut {NoStop}%
\bibitem [{\citenamefont {Levi}\ \emph {et~al.}(2012)\citenamefont {Levi},
  \citenamefont {Krivolapov}, \citenamefont {Fishman},\ and\ \citenamefont
  {Segev}}]{LeviEtAl2012}%
  \BibitemOpen
  \bibfield  {author} {\bibinfo {author} {\bibfnamefont {L.}~\bibnamefont
  {Levi}}, \bibinfo {author} {\bibfnamefont {Y.}~\bibnamefont {Krivolapov}},
  \bibinfo {author} {\bibfnamefont {S.}~\bibnamefont {Fishman}}, \ and\
  \bibinfo {author} {\bibfnamefont {M.}~\bibnamefont {Segev}},\ }\bibfield
  {title} {\enquote {\bibinfo {title} {Hyper-transport of light and stochastic
  acceleration by evolving disorder},}\ }\href {\doibase 10.1038/nphys2463}
  {\bibfield  {journal} {\bibinfo  {journal} {Nat. Phys.}\ }\textbf {\bibinfo
  {volume} {8}},\ \bibinfo {pages} {912} (\bibinfo {year} {2012})}\BibitemShut
  {NoStop}%
\bibitem [{\citenamefont {Hufnagel}\ \emph {et~al.}(2001)\citenamefont
  {Hufnagel}, \citenamefont {Ketzmerick}, \citenamefont {Kottos},\ and\
  \citenamefont {Geisel}}]{HufnagelEtAl2001}%
  \BibitemOpen
  \bibfield  {author} {\bibinfo {author} {\bibfnamefont {L.}~\bibnamefont
  {Hufnagel}}, \bibinfo {author} {\bibfnamefont {R.}~\bibnamefont
  {Ketzmerick}}, \bibinfo {author} {\bibfnamefont {T.}~\bibnamefont {Kottos}},
  \ and\ \bibinfo {author} {\bibfnamefont {T.}~\bibnamefont {Geisel}},\
  }\bibfield  {title} {\enquote {\bibinfo {title} {Superballistic spreading of
  wave packets},}\ }\href {\doibase 10.1103/PhysRevE.64.012301} {\bibfield
  {journal} {\bibinfo  {journal} {Phys. Rev. E}\ }\textbf {\bibinfo {volume}
  {64}},\ \bibinfo {pages} {012301} (\bibinfo {year} {2001})}\BibitemShut
  {NoStop}%
\bibitem [{\citenamefont {Zhang}\ \emph {et~al.}(2012)\citenamefont {Zhang},
  \citenamefont {Mao},\ and\ \citenamefont {Zhang}}]{ZhangEtAl2012}%
  \BibitemOpen
  \bibfield  {author} {\bibinfo {author} {\bibfnamefont {Y.}~\bibnamefont
  {Zhang}}, \bibinfo {author} {\bibfnamefont {L.}~\bibnamefont {Mao}}, \ and\
  \bibinfo {author} {\bibfnamefont {C.}~\bibnamefont {Zhang}},\ }\bibfield
  {title} {\enquote {\bibinfo {title} {Mean-field dynamics of spin-orbit
  coupled {Bose-Einstein} condensates},}\ }\href {\doibase
  10.1103/PhysRevLett.108.035302} {\bibfield  {journal} {\bibinfo  {journal}
  {Phys. Rev. Lett.}\ }\textbf {\bibinfo {volume} {108}},\ \bibinfo {pages}
  {035302} (\bibinfo {year} {2012})}\BibitemShut {NoStop}%
\bibitem [{\citenamefont {St\"{u}tzer}\ \emph {et~al.}(2013)\citenamefont
  {St\"{u}tzer}, \citenamefont {Kottos}, \citenamefont {T\"{u}nnermann},
  \citenamefont {Nolte}, \citenamefont {Christodoulides},\ and\ \citenamefont
  {Szameit}}]{StutzerEtAl2013}%
  \BibitemOpen
  \bibfield  {author} {\bibinfo {author} {\bibfnamefont {S.}~\bibnamefont
  {St\"{u}tzer}}, \bibinfo {author} {\bibfnamefont {T.}~\bibnamefont {Kottos}},
  \bibinfo {author} {\bibfnamefont {A.}~\bibnamefont {T\"{u}nnermann}},
  \bibinfo {author} {\bibfnamefont {S.}~\bibnamefont {Nolte}}, \bibinfo
  {author} {\bibfnamefont {D.~N.}\ \bibnamefont {Christodoulides}}, \ and\
  \bibinfo {author} {\bibfnamefont {A.}~\bibnamefont {Szameit}},\ }\bibfield
  {title} {\enquote {\bibinfo {title} {Superballistic growth of the variance of
  optical wave packets},}\ }\href {\doibase 10.1364/OL.38.004675} {\bibfield
  {journal} {\bibinfo  {journal} {Opt. Lett.}\ }\textbf {\bibinfo {volume}
  {38}},\ \bibinfo {pages} {4675} (\bibinfo {year} {2013})}\BibitemShut
  {NoStop}%
\bibitem [{\citenamefont {Zhao}\ \emph {et~al.}(2014)\citenamefont {Zhao},
  \citenamefont {M\"uller},\ and\ \citenamefont {Gong}}]{ZhaoEtAl2014}%
  \BibitemOpen
  \bibfield  {author} {\bibinfo {author} {\bibfnamefont {Q.}~\bibnamefont
  {Zhao}}, \bibinfo {author} {\bibfnamefont {C.~A.}\ \bibnamefont {M\"uller}},
  \ and\ \bibinfo {author} {\bibfnamefont {J.}~\bibnamefont {Gong}},\
  }\bibfield  {title} {\enquote {\bibinfo {title} {Quantum and classical
  superballistic transport in a relativistic kicked-rotor system},}\ }\href
  {\doibase 10.1103/PhysRevE.90.022921} {\bibfield  {journal} {\bibinfo
  {journal} {Phys. Rev. E}\ }\textbf {\bibinfo {volume} {90}},\ \bibinfo
  {pages} {022921} (\bibinfo {year} {2014})}\BibitemShut {NoStop}%
\bibitem [{\citenamefont {Weiss}\ \emph
  {et~al.}(2015{\natexlab{a}})\citenamefont {Weiss}, \citenamefont {Gardiner},\
  and\ \citenamefont {Breuer}}]{WeissEtAl2015}%
  \BibitemOpen
  \bibfield  {author} {\bibinfo {author} {\bibfnamefont {C.}~\bibnamefont
  {Weiss}}, \bibinfo {author} {\bibfnamefont {S.~A.}\ \bibnamefont {Gardiner}},
  \ and\ \bibinfo {author} {\bibfnamefont {H.-P.}\ \bibnamefont {Breuer}},\
  }\bibfield  {title} {\enquote {\bibinfo {title} {From short-time diffusive to
  long-time ballistic dynamics: The unusual center-of-mass motion of quantum
  bright solitons},}\ }\href {\doibase 10.1103/PhysRevA.91.063616} {\bibfield
  {journal} {\bibinfo  {journal} {Phys. Rev. A}\ }\textbf {\bibinfo {volume}
  {91}},\ \bibinfo {pages} {063616} (\bibinfo {year}
  {2015}{\natexlab{a}})}\BibitemShut {NoStop}%
\bibitem [{\citenamefont {Khaykovich}\ \emph {et~al.}(2002)\citenamefont
  {Khaykovich}, \citenamefont {Schreck}, \citenamefont {Ferrari}, \citenamefont
  {Bourdel}, \citenamefont {Cubizolles}, \citenamefont {Carr}, \citenamefont
  {Castin},\ and\ \citenamefont {Salomon}}]{KhaykovichEtAl2002}%
  \BibitemOpen
  \bibfield  {author} {\bibinfo {author} {\bibfnamefont {L.}~\bibnamefont
  {Khaykovich}}, \bibinfo {author} {\bibfnamefont {F.}~\bibnamefont {Schreck}},
  \bibinfo {author} {\bibfnamefont {G.}~\bibnamefont {Ferrari}}, \bibinfo
  {author} {\bibfnamefont {T.}~\bibnamefont {Bourdel}}, \bibinfo {author}
  {\bibfnamefont {J.}~\bibnamefont {Cubizolles}}, \bibinfo {author}
  {\bibfnamefont {L.~D.}\ \bibnamefont {Carr}}, \bibinfo {author}
  {\bibfnamefont {Y.}~\bibnamefont {Castin}}, \ and\ \bibinfo {author}
  {\bibfnamefont {C.}~\bibnamefont {Salomon}},\ }\bibfield  {title} {\enquote
  {\bibinfo {title} {Formation of a matter-wave bright soliton},}\ }\href
  {\doibase 10.1126/science.1071021} {\bibfield  {journal} {\bibinfo  {journal}
  {Science}\ }\textbf {\bibinfo {volume} {296}},\ \bibinfo {pages} {1290}
  (\bibinfo {year} {2002})}\BibitemShut {NoStop}%
\bibitem [{\citenamefont {Strecker}\ \emph {et~al.}(2002)\citenamefont
  {Strecker}, \citenamefont {Partridge}, \citenamefont {Truscott},\ and\
  \citenamefont {Hulet}}]{StreckerEtAl2002}%
  \BibitemOpen
  \bibfield  {author} {\bibinfo {author} {\bibfnamefont {K.~E.}\ \bibnamefont
  {Strecker}}, \bibinfo {author} {\bibfnamefont {G.~B.}\ \bibnamefont
  {Partridge}}, \bibinfo {author} {\bibfnamefont {A.~G.}\ \bibnamefont
  {Truscott}}, \ and\ \bibinfo {author} {\bibfnamefont {R.~G.}\ \bibnamefont
  {Hulet}},\ }\bibfield  {title} {\enquote {\bibinfo {title} {Formation and
  propagation of matter-wave soliton trains},}\ }\href {\doibase
  10.1038/nature747} {\bibfield  {journal} {\bibinfo  {journal} {Nature
  (London)}\ }\textbf {\bibinfo {volume} {417}},\ \bibinfo {pages} {150}
  (\bibinfo {year} {2002})}\BibitemShut {NoStop}%
\bibitem [{\citenamefont {Cornish}\ \emph {et~al.}(2006)\citenamefont
  {Cornish}, \citenamefont {Thompson},\ and\ \citenamefont
  {Wieman}}]{CornishEtAl2006}%
  \BibitemOpen
  \bibfield  {author} {\bibinfo {author} {\bibfnamefont {S.~L.}\ \bibnamefont
  {Cornish}}, \bibinfo {author} {\bibfnamefont {S.~T.}\ \bibnamefont
  {Thompson}}, \ and\ \bibinfo {author} {\bibfnamefont {C.~E.}\ \bibnamefont
  {Wieman}},\ }\bibfield  {title} {\enquote {\bibinfo {title} {Formation of
  bright matter-wave solitons during the collapse of attractive {Bose-Einstein}
  condensates},}\ }\href {\doibase 10.1103/PhysRevLett.96.170401} {\bibfield
  {journal} {\bibinfo  {journal} {Phys. Rev. Lett.}\ }\textbf {\bibinfo
  {volume} {96}},\ \bibinfo {pages} {170401} (\bibinfo {year}
  {2006})}\BibitemShut {NoStop}%
\bibitem [{\citenamefont {{Marchant}}\ \emph {et~al.}(2013)\citenamefont
  {{Marchant}}, \citenamefont {{Billam}}, \citenamefont {{Wiles}},
  \citenamefont {{Yu}}, \citenamefont {{Gardiner}},\ and\ \citenamefont
  {{Cornish}}}]{MarchantEtAl2013}%
  \BibitemOpen
  \bibfield  {author} {\bibinfo {author} {\bibfnamefont {A.~L.}\ \bibnamefont
  {{Marchant}}}, \bibinfo {author} {\bibfnamefont {T.~P.}\ \bibnamefont
  {{Billam}}}, \bibinfo {author} {\bibfnamefont {T.~P.}\ \bibnamefont
  {{Wiles}}}, \bibinfo {author} {\bibfnamefont {M.~M.~H.}\ \bibnamefont
  {{Yu}}}, \bibinfo {author} {\bibfnamefont {S.~A.}\ \bibnamefont
  {{Gardiner}}}, \ and\ \bibinfo {author} {\bibfnamefont {S.~L.}\ \bibnamefont
  {{Cornish}}},\ }\bibfield  {title} {\enquote {\bibinfo {title} {{Controlled
  formation and reflection of a bright solitary matter-wave}},}\ }\href
  {\doibase 10.1038/ncomms2893} {\bibfield  {journal} {\bibinfo  {journal}
  {Nat. Commun.}\ }\textbf {\bibinfo {volume} {4}},\ \bibinfo {pages} {1865}
  (\bibinfo {year} {2013})}\BibitemShut {NoStop}%
\bibitem [{\citenamefont {Medley}\ \emph {et~al.}(2014)\citenamefont {Medley},
  \citenamefont {Minar}, \citenamefont {Cizek}, \citenamefont {Berryrieser},\
  and\ \citenamefont {Kasevich}}]{MedleyEtAl2014}%
  \BibitemOpen
  \bibfield  {author} {\bibinfo {author} {\bibfnamefont {P.}~\bibnamefont
  {Medley}}, \bibinfo {author} {\bibfnamefont {M.~A.}\ \bibnamefont {Minar}},
  \bibinfo {author} {\bibfnamefont {N.~C.}\ \bibnamefont {Cizek}}, \bibinfo
  {author} {\bibfnamefont {D.}~\bibnamefont {Berryrieser}}, \ and\ \bibinfo
  {author} {\bibfnamefont {M.~A.}\ \bibnamefont {Kasevich}},\ }\bibfield
  {title} {\enquote {\bibinfo {title} {Evaporative production of bright atomic
  solitons},}\ }\href {\doibase 10.1103/PhysRevLett.112.060401} {\bibfield
  {journal} {\bibinfo  {journal} {Phys. Rev. Lett.}\ }\textbf {\bibinfo
  {volume} {112}},\ \bibinfo {pages} {060401} (\bibinfo {year}
  {2014})}\BibitemShut {NoStop}%
\bibitem [{\citenamefont {McDonald}\ \emph {et~al.}(2014)\citenamefont
  {McDonald}, \citenamefont {Kuhn}, \citenamefont {Hardman}, \citenamefont
  {Bennetts}, \citenamefont {Everitt}, \citenamefont {Altin}, \citenamefont
  {Debs}, \citenamefont {Close},\ and\ \citenamefont
  {Robins}}]{McDonaldEtAl2014}%
  \BibitemOpen
  \bibfield  {author} {\bibinfo {author} {\bibfnamefont {G.~D.}\ \bibnamefont
  {McDonald}}, \bibinfo {author} {\bibfnamefont {C.~C.~N.}\ \bibnamefont
  {Kuhn}}, \bibinfo {author} {\bibfnamefont {K.~S.}\ \bibnamefont {Hardman}},
  \bibinfo {author} {\bibfnamefont {S.}~\bibnamefont {Bennetts}}, \bibinfo
  {author} {\bibfnamefont {P.~J.}\ \bibnamefont {Everitt}}, \bibinfo {author}
  {\bibfnamefont {P.~A.}\ \bibnamefont {Altin}}, \bibinfo {author}
  {\bibfnamefont {J.~E.}\ \bibnamefont {Debs}}, \bibinfo {author}
  {\bibfnamefont {J.~D.}\ \bibnamefont {Close}}, \ and\ \bibinfo {author}
  {\bibfnamefont {N.~P.}\ \bibnamefont {Robins}},\ }\bibfield  {title}
  {\enquote {\bibinfo {title} {Bright solitonic matter-wave interferometer},}\
  }\href {\doibase 10.1103/PhysRevLett.113.013002} {\bibfield  {journal}
  {\bibinfo  {journal} {Phys. Rev. Lett.}\ }\textbf {\bibinfo {volume} {113}},\
  \bibinfo {pages} {013002} (\bibinfo {year} {2014})}\BibitemShut {NoStop}%
\bibitem [{\citenamefont {{Nguyen}}\ \emph {et~al.}(2014)\citenamefont
  {{Nguyen}}, \citenamefont {{Dyke}}, \citenamefont {{Luo}}, \citenamefont
  {{Malomed}},\ and\ \citenamefont {{Hulet}}}]{NguyenEtAl2014}%
  \BibitemOpen
  \bibfield  {author} {\bibinfo {author} {\bibfnamefont {J.~H.~V.}\
  \bibnamefont {{Nguyen}}}, \bibinfo {author} {\bibfnamefont {P.}~\bibnamefont
  {{Dyke}}}, \bibinfo {author} {\bibfnamefont {D.}~\bibnamefont {{Luo}}},
  \bibinfo {author} {\bibfnamefont {B.~A.}\ \bibnamefont {{Malomed}}}, \ and\
  \bibinfo {author} {\bibfnamefont {R.~G.}\ \bibnamefont {{Hulet}}},\
  }\bibfield  {title} {\enquote {\bibinfo {title} {{Collisions of matter-wave
  solitons}},}\ }\href {\doibase 10.1038/nphys3135} {\bibfield  {journal}
  {\bibinfo  {journal} {Nat. Phys.}\ }\textbf {\bibinfo {volume} {10}},\
  \bibinfo {pages} {918} (\bibinfo {year} {2014})}\BibitemShut {NoStop}%
\bibitem [{\citenamefont {Marchant}\ \emph {et~al.}(2016)\citenamefont
  {Marchant}, \citenamefont {Billam}, \citenamefont {Yu}, \citenamefont
  {Rakonjac}, \citenamefont {Helm}, \citenamefont {Polo}, \citenamefont
  {Weiss}, \citenamefont {Gardiner},\ and\ \citenamefont
  {Cornish}}]{MarchantEtAl2016}%
  \BibitemOpen
  \bibfield  {author} {\bibinfo {author} {\bibfnamefont {A.~L.}\ \bibnamefont
  {Marchant}}, \bibinfo {author} {\bibfnamefont {T.~P.}\ \bibnamefont
  {Billam}}, \bibinfo {author} {\bibfnamefont {M.~M.~H.}\ \bibnamefont {Yu}},
  \bibinfo {author} {\bibfnamefont {A.}~\bibnamefont {Rakonjac}}, \bibinfo
  {author} {\bibfnamefont {J.~L.}\ \bibnamefont {Helm}}, \bibinfo {author}
  {\bibfnamefont {J.}~\bibnamefont {Polo}}, \bibinfo {author} {\bibfnamefont
  {C.}~\bibnamefont {Weiss}}, \bibinfo {author} {\bibfnamefont {S.~A.}\
  \bibnamefont {Gardiner}}, \ and\ \bibinfo {author} {\bibfnamefont {S.~L.}\
  \bibnamefont {Cornish}},\ }\bibfield  {title} {\enquote {\bibinfo {title}
  {Quantum reflection of bright solitary matter waves from a narrow attractive
  potential},}\ }\href {\doibase 10.1103/PhysRevA.93.021604} {\bibfield
  {journal} {\bibinfo  {journal} {Phys. Rev. A}\ }\textbf {\bibinfo {volume}
  {93}},\ \bibinfo {pages} {021604(R)} (\bibinfo {year} {2016})}\BibitemShut
  {NoStop}%
\bibitem [{\citenamefont {{Everitt}}\ \emph {et~al.}(2015)\citenamefont
  {{Everitt}}, \citenamefont {{Sooriyabandara}}, \citenamefont {{McDonald}},
  \citenamefont {{Hardman}}, \citenamefont {{Quinlivan}}, \citenamefont
  {{Perumbil}}, \citenamefont {{Wigley}}, \citenamefont {{Debs}}, \citenamefont
  {{Close}}, \citenamefont {{Kuhn}},\ and\ \citenamefont
  {{Robins}}}]{EverittEtAl2015}%
  \BibitemOpen
  \bibfield  {author} {\bibinfo {author} {\bibfnamefont {P.~J.}\ \bibnamefont
  {{Everitt}}}, \bibinfo {author} {\bibfnamefont {M.~A.}\ \bibnamefont
  {{Sooriyabandara}}}, \bibinfo {author} {\bibfnamefont {G.~D.}\ \bibnamefont
  {{McDonald}}}, \bibinfo {author} {\bibfnamefont {K.~S.}\ \bibnamefont
  {{Hardman}}}, \bibinfo {author} {\bibfnamefont {C.}~\bibnamefont
  {{Quinlivan}}}, \bibinfo {author} {\bibfnamefont {M.}~\bibnamefont
  {{Perumbil}}}, \bibinfo {author} {\bibfnamefont {P.}~\bibnamefont
  {{Wigley}}}, \bibinfo {author} {\bibfnamefont {J.~E.}\ \bibnamefont
  {{Debs}}}, \bibinfo {author} {\bibfnamefont {J.~D.}\ \bibnamefont {{Close}}},
  \bibinfo {author} {\bibfnamefont {C.~C.~N.}\ \bibnamefont {{Kuhn}}}, \ and\
  \bibinfo {author} {\bibfnamefont {N.~P.}\ \bibnamefont {{Robins}}},\
  }\bibfield  {title} {\enquote {\bibinfo {title} {{Observation of Breathers in
  an Attractive Bose Gas}},}\ }\href@noop {} {\bibfield  {journal} {\bibinfo
  {journal} {ArXiv e-prints}\ } (\bibinfo {year} {2015})},\ \Eprint
  {http://arxiv.org/abs/1509.06844} {arXiv:1509.06844 [cond-mat.quant-gas]}
  \BibitemShut {NoStop}%
\bibitem [{\citenamefont {Pethick}\ and\ \citenamefont
  {Smith}(2008)}]{PethickSmith2008}%
  \BibitemOpen
  \bibfield  {author} {\bibinfo {author} {\bibfnamefont {C.~J.}\ \bibnamefont
  {Pethick}}\ and\ \bibinfo {author} {\bibfnamefont {H.}~\bibnamefont
  {Smith}},\ }\href@noop {} {\emph {\bibinfo {title} {Bose-Einstein
  Condensation in Dilute Gases}}}\ (\bibinfo  {publisher} {Cambridge University
  Press},\ \bibinfo {address} {Cambridge},\ \bibinfo {year} {2008})\BibitemShut
  {NoStop}%
\bibitem [{\citenamefont {Baizakov}\ \emph {et~al.}(2002)\citenamefont
  {Baizakov}, \citenamefont {Konotop},\ and\ \citenamefont
  {Salerno}}]{BaizakovEtAl2002}%
  \BibitemOpen
  \bibfield  {author} {\bibinfo {author} {\bibfnamefont {B.~B.}\ \bibnamefont
  {Baizakov}}, \bibinfo {author} {\bibfnamefont {V.~V.}\ \bibnamefont
  {Konotop}}, \ and\ \bibinfo {author} {\bibfnamefont {M.}~\bibnamefont
  {Salerno}},\ }\bibfield  {title} {\enquote {\bibinfo {title} {Regular spatial
  structures in arrays of {Bose-Einstein} condensates induced by modulational
  instability},}\ }\href {\doibase 10.1088/0953-4075/35/24/312} {\bibfield
  {journal} {\bibinfo  {journal} {J. Phys. B}\ }\textbf {\bibinfo {volume}
  {35}},\ \bibinfo {pages} {5105} (\bibinfo {year} {2002})}\BibitemShut
  {NoStop}%
\bibitem [{\citenamefont {Al~Khawaja}\ \emph {et~al.}(2002)\citenamefont
  {Al~Khawaja}, \citenamefont {Stoof}, \citenamefont {Hulet}, \citenamefont
  {Strecker},\ and\ \citenamefont {Partridge}}]{AlKhawajaEtAl2002}%
  \BibitemOpen
  \bibfield  {author} {\bibinfo {author} {\bibfnamefont {U.}~\bibnamefont
  {Al~Khawaja}}, \bibinfo {author} {\bibfnamefont {H.~T.~C.}\ \bibnamefont
  {Stoof}}, \bibinfo {author} {\bibfnamefont {R.~G.}\ \bibnamefont {Hulet}},
  \bibinfo {author} {\bibfnamefont {K.~E.}\ \bibnamefont {Strecker}}, \ and\
  \bibinfo {author} {\bibfnamefont {G.~B.}\ \bibnamefont {Partridge}},\
  }\bibfield  {title} {\enquote {\bibinfo {title} {Bright soliton trains of
  trapped {Bose-Einstein} condensates},}\ }\href {\doibase
  10.1103/PhysRevLett.89.200404} {\bibfield  {journal} {\bibinfo  {journal}
  {Phys. Rev. Lett.}\ }\textbf {\bibinfo {volume} {89}},\ \bibinfo {pages}
  {200404} (\bibinfo {year} {2002})}\BibitemShut {NoStop}%
\bibitem [{\citenamefont {Hai}\ \emph {et~al.}(2004)\citenamefont {Hai},
  \citenamefont {Lee},\ and\ \citenamefont {Chong}}]{HaiEtAl2004}%
  \BibitemOpen
  \bibfield  {author} {\bibinfo {author} {\bibfnamefont {W.}~\bibnamefont
  {Hai}}, \bibinfo {author} {\bibfnamefont {C.}~\bibnamefont {Lee}}, \ and\
  \bibinfo {author} {\bibfnamefont {G.}~\bibnamefont {Chong}},\ }\bibfield
  {title} {\enquote {\bibinfo {title} {Propagation and breathing of
  matter-wave-packet trains},}\ }\href {\doibase 10.1103/PhysRevA.70.053621}
  {\bibfield  {journal} {\bibinfo  {journal} {Phys. Rev. A}\ }\textbf {\bibinfo
  {volume} {70}},\ \bibinfo {pages} {053621} (\bibinfo {year}
  {2004})}\BibitemShut {NoStop}%
\bibitem [{\citenamefont {Martin}\ and\ \citenamefont
  {Ruostekoski}(2012)}]{MartinRuostekoski2012}%
  \BibitemOpen
  \bibfield  {author} {\bibinfo {author} {\bibfnamefont {A.~D.}\ \bibnamefont
  {Martin}}\ and\ \bibinfo {author} {\bibfnamefont {J.}~\bibnamefont
  {Ruostekoski}},\ }\bibfield  {title} {\enquote {\bibinfo {title} {Quantum
  dynamics of atomic bright solitons under splitting and recollision, and
  implications for interferometry},}\ }\href {\doibase
  10.1088/1367-2630/14/4/043040} {\bibfield  {journal} {\bibinfo  {journal}
  {New J. Phys.}\ }\textbf {\bibinfo {volume} {14}},\ \bibinfo {pages} {043040}
  (\bibinfo {year} {2012})}\BibitemShut {NoStop}%
\bibitem [{\citenamefont {Cuevas}\ \emph {et~al.}(2013)\citenamefont {Cuevas},
  \citenamefont {Kevrekidis}, \citenamefont {Malomed}, \citenamefont {Dyke},\
  and\ \citenamefont {Hulet}}]{CuevasEtAl2013}%
  \BibitemOpen
  \bibfield  {author} {\bibinfo {author} {\bibfnamefont {J.}~\bibnamefont
  {Cuevas}}, \bibinfo {author} {\bibfnamefont {P.~G.}\ \bibnamefont
  {Kevrekidis}}, \bibinfo {author} {\bibfnamefont {B.~A.}\ \bibnamefont
  {Malomed}}, \bibinfo {author} {\bibfnamefont {P.}~\bibnamefont {Dyke}}, \
  and\ \bibinfo {author} {\bibfnamefont {R.~G.}\ \bibnamefont {Hulet}},\
  }\bibfield  {title} {\enquote {\bibinfo {title} {Interactions of solitons
  with a {Gaussian} barrier: splitting and recombination in
  quasi-one-dimensional and three-dimensional settings},}\ }\href {\doibase
  10.1088/1367-2630/15/6/063006} {\bibfield  {journal} {\bibinfo  {journal}
  {New J. Phys.}\ }\textbf {\bibinfo {volume} {15}},\ \bibinfo {pages} {063006}
  (\bibinfo {year} {2013})}\BibitemShut {NoStop}%
\bibitem [{\citenamefont {Polo}\ and\ \citenamefont
  {Ahufinger}(2013)}]{PoloAhufinger2013}%
  \BibitemOpen
  \bibfield  {author} {\bibinfo {author} {\bibfnamefont {J.}~\bibnamefont
  {Polo}}\ and\ \bibinfo {author} {\bibfnamefont {V.}~\bibnamefont
  {Ahufinger}},\ }\bibfield  {title} {\enquote {\bibinfo {title} {Soliton-based
  matter-wave interferometer},}\ }\href {\doibase 10.1103/PhysRevA.88.053628}
  {\bibfield  {journal} {\bibinfo  {journal} {Phys. Rev. A}\ }\textbf {\bibinfo
  {volume} {88}},\ \bibinfo {pages} {053628} (\bibinfo {year}
  {2013})}\BibitemShut {NoStop}%
\bibitem [{\citenamefont {Sun}\ \emph {et~al.}(2014)\citenamefont {Sun},
  \citenamefont {Kevrekidis},\ and\ \citenamefont {Kr\"uger}}]{SunEtAl2014}%
  \BibitemOpen
  \bibfield  {author} {\bibinfo {author} {\bibfnamefont {Z.-Y.}\ \bibnamefont
  {Sun}}, \bibinfo {author} {\bibfnamefont {P.~G.}\ \bibnamefont {Kevrekidis}},
  \ and\ \bibinfo {author} {\bibfnamefont {P.}~\bibnamefont {Kr\"uger}},\
  }\bibfield  {title} {\enquote {\bibinfo {title} {Mean-field analog of the
  {Hong-Ou-Mandel} experiment with bright solitons},}\ }\href {\doibase
  10.1103/PhysRevA.90.063612} {\bibfield  {journal} {\bibinfo  {journal} {Phys.
  Rev. A}\ }\textbf {\bibinfo {volume} {90}},\ \bibinfo {pages} {063612}
  (\bibinfo {year} {2014})}\BibitemShut {NoStop}%
\bibitem [{\citenamefont {Helm}\ \emph {et~al.}(2015)\citenamefont {Helm},
  \citenamefont {Cornish},\ and\ \citenamefont {Gardiner}}]{HelmEtAl2015}%
  \BibitemOpen
  \bibfield  {author} {\bibinfo {author} {\bibfnamefont {J.~L.}\ \bibnamefont
  {Helm}}, \bibinfo {author} {\bibfnamefont {S.~L.}\ \bibnamefont {Cornish}}, \
  and\ \bibinfo {author} {\bibfnamefont {S.~A.}\ \bibnamefont {Gardiner}},\
  }\bibfield  {title} {\enquote {\bibinfo {title} {Sagnac interferometry using
  bright matter-wave solitons},}\ }\href {\doibase
  10.1103/PhysRevLett.114.134101} {\bibfield  {journal} {\bibinfo  {journal}
  {Phys. Rev. Lett.}\ }\textbf {\bibinfo {volume} {114}},\ \bibinfo {pages}
  {134101} (\bibinfo {year} {2015})}\BibitemShut {NoStop}%
\bibitem [{\citenamefont {{Dunjko}}\ and\ \citenamefont
  {{Olshanii}}(2015)}]{DunjkoOlshanii2015}%
  \BibitemOpen
  \bibfield  {author} {\bibinfo {author} {\bibfnamefont {V.}~\bibnamefont
  {{Dunjko}}}\ and\ \bibinfo {author} {\bibfnamefont {M.}~\bibnamefont
  {{Olshanii}}},\ }\bibfield  {title} {\enquote {\bibinfo {title} {{Superheated
  integrability and multisoliton survival through scattering off barriers}},}\
  }\href@noop {} {\bibfield  {journal} {\bibinfo  {journal} {ArXiv e-prints}\ }
  (\bibinfo {year} {2015})},\ \Eprint {http://arxiv.org/abs/1501.00075}
  {arXiv:1501.00075 [cond-mat.quant-gas]} \BibitemShut {NoStop}%
\bibitem [{\citenamefont {Lai}\ and\ \citenamefont {Haus}(1989)}]{LaiHaus1989}%
  \BibitemOpen
  \bibfield  {author} {\bibinfo {author} {\bibfnamefont {Y.}~\bibnamefont
  {Lai}}\ and\ \bibinfo {author} {\bibfnamefont {H.~A.}\ \bibnamefont {Haus}},\
  }\bibfield  {title} {\enquote {\bibinfo {title} {Quantum theory of solitons
  in optical fibers. ii. exact solution},}\ }\href {\doibase
  10.1103/PhysRevA.40.854} {\bibfield  {journal} {\bibinfo  {journal} {Phys.
  Rev. A}\ }\textbf {\bibinfo {volume} {40}},\ \bibinfo {pages} {854} (\bibinfo
  {year} {1989})}\BibitemShut {NoStop}%
\bibitem [{\citenamefont {{Drummond}}\ \emph {et~al.}(1993)\citenamefont
  {{Drummond}}, \citenamefont {{Shelby}}, \citenamefont {{Friberg}},\ and\
  \citenamefont {{Yamamoto}}}]{DrummondEtAl1993}%
  \BibitemOpen
  \bibfield  {author} {\bibinfo {author} {\bibfnamefont {P.~D.}\ \bibnamefont
  {{Drummond}}}, \bibinfo {author} {\bibfnamefont {R.~M.}\ \bibnamefont
  {{Shelby}}}, \bibinfo {author} {\bibfnamefont {S.~R.}\ \bibnamefont
  {{Friberg}}}, \ and\ \bibinfo {author} {\bibfnamefont {Y.}~\bibnamefont
  {{Yamamoto}}},\ }\bibfield  {title} {\enquote {\bibinfo {title} {{Quantum
  solitons in optical fibres}},}\ }\href {\doibase 10.1038/365307a0} {\bibfield
   {journal} {\bibinfo  {journal} {Nature (London)}\ }\textbf {\bibinfo
  {volume} {365}},\ \bibinfo {pages} {307} (\bibinfo {year}
  {1993})}\BibitemShut {NoStop}%
\bibitem [{\citenamefont {Carr}\ and\ \citenamefont
  {Brand}(2004)}]{CarrBrand2004}%
  \BibitemOpen
  \bibfield  {author} {\bibinfo {author} {\bibfnamefont {L.~D.}\ \bibnamefont
  {Carr}}\ and\ \bibinfo {author} {\bibfnamefont {J.}~\bibnamefont {Brand}},\
  }\bibfield  {title} {\enquote {\bibinfo {title} {Spontaneous soliton
  formation and modulational instability in {Bose-Einstein} condensates},}\
  }\href {\doibase 10.1103/PhysRevLett.92.040401} {\bibfield  {journal}
  {\bibinfo  {journal} {Phys. Rev. Lett.}\ }\textbf {\bibinfo {volume} {92}},\
  \bibinfo {pages} {040401} (\bibinfo {year} {2004})}\BibitemShut {NoStop}%
\bibitem [{\citenamefont {Mishmash}\ and\ \citenamefont
  {Carr}(2009)}]{MishmashCarr2009}%
  \BibitemOpen
  \bibfield  {author} {\bibinfo {author} {\bibfnamefont {R.~V.}\ \bibnamefont
  {Mishmash}}\ and\ \bibinfo {author} {\bibfnamefont {L.~D.}\ \bibnamefont
  {Carr}},\ }\bibfield  {title} {\enquote {\bibinfo {title} {Quantum entangled
  dark solitons formed by ultracold atoms in optical lattices},}\ }\href
  {\doibase 10.1103/PhysRevLett.103.140403} {\bibfield  {journal} {\bibinfo
  {journal} {Phys. Rev. Lett.}\ }\textbf {\bibinfo {volume} {103}},\ \bibinfo
  {pages} {140403} (\bibinfo {year} {2009})}\BibitemShut {NoStop}%
\bibitem [{\citenamefont {Streltsov}\ \emph {et~al.}(2011)\citenamefont
  {Streltsov}, \citenamefont {Alon},\ and\ \citenamefont
  {Cederbaum}}]{StreltsovEtAl2011}%
  \BibitemOpen
  \bibfield  {author} {\bibinfo {author} {\bibfnamefont {A.~I.}\ \bibnamefont
  {Streltsov}}, \bibinfo {author} {\bibfnamefont {O.~E.}\ \bibnamefont {Alon}},
  \ and\ \bibinfo {author} {\bibfnamefont {L.~S.}\ \bibnamefont {Cederbaum}},\
  }\bibfield  {title} {\enquote {\bibinfo {title} {Swift loss of coherence of
  soliton trains in attractive {Bose-Einstein} condensates},}\ }\href {\doibase
  10.1103/PhysRevLett.106.240401} {\bibfield  {journal} {\bibinfo  {journal}
  {Phys. Rev. Lett.}\ }\textbf {\bibinfo {volume} {106}},\ \bibinfo {pages}
  {240401} (\bibinfo {year} {2011})}\BibitemShut {NoStop}%
\bibitem [{\citenamefont {{Fogarty}}\ \emph {et~al.}(2013)\citenamefont
  {{Fogarty}}, \citenamefont {{Kiely}}, \citenamefont {{Campbell}},\ and\
  \citenamefont {{Busch}}}]{FogartyEtAl2013}%
  \BibitemOpen
  \bibfield  {author} {\bibinfo {author} {\bibfnamefont {T.}~\bibnamefont
  {{Fogarty}}}, \bibinfo {author} {\bibfnamefont {A.}~\bibnamefont {{Kiely}}},
  \bibinfo {author} {\bibfnamefont {S.}~\bibnamefont {{Campbell}}}, \ and\
  \bibinfo {author} {\bibfnamefont {T.}~\bibnamefont {{Busch}}},\ }\bibfield
  {title} {\enquote {\bibinfo {title} {{Effect of interparticle interaction in
  a free-oscillation atomic interferometer}},}\ }\href {\doibase
  10.1103/PhysRevA.87.043630} {\bibfield  {journal} {\bibinfo  {journal}
  {Phys.\ Rev.\ A}\ }\textbf {\bibinfo {volume} {87}},\ \bibinfo {eid} {043630}
  (\bibinfo {year} {2013})}\BibitemShut {NoStop}%
\bibitem [{\citenamefont {{Delande}}\ \emph {et~al.}(2013)\citenamefont
  {{Delande}}, \citenamefont {{Sacha}}, \citenamefont {{P{\l}odzie{\'n}}},
  \citenamefont {{Avazbaev}},\ and\ \citenamefont
  {{Zakrzewski}}}]{DelandeEtAl2013}%
  \BibitemOpen
  \bibfield  {author} {\bibinfo {author} {\bibfnamefont {D.}~\bibnamefont
  {{Delande}}}, \bibinfo {author} {\bibfnamefont {K.}~\bibnamefont {{Sacha}}},
  \bibinfo {author} {\bibfnamefont {M.}~\bibnamefont {{P{\l}odzie{\'n}}}},
  \bibinfo {author} {\bibfnamefont {S.~K.}\ \bibnamefont {{Avazbaev}}}, \ and\
  \bibinfo {author} {\bibfnamefont {J.}~\bibnamefont {{Zakrzewski}}},\
  }\bibfield  {title} {\enquote {\bibinfo {title} {{Many-body Anderson
  localization in one-dimensional systems}},}\ }\href {\doibase
  10.1088/1367-2630/15/4/045021} {\bibfield  {journal} {\bibinfo  {journal}
  {New J. Phys.}\ }\textbf {\bibinfo {volume} {15}},\ \bibinfo {eid} {045021}
  (\bibinfo {year} {2013})}\BibitemShut {NoStop}%
\bibitem [{\citenamefont {Gertjerenken}\ \emph {et~al.}(2013)\citenamefont
  {Gertjerenken}, \citenamefont {Billam}, \citenamefont {Blackley},
  \citenamefont {Le~Sueur}, \citenamefont {Khaykovich}, \citenamefont
  {Cornish},\ and\ \citenamefont {Weiss}}]{GertjerenkenEtAl2013}%
  \BibitemOpen
  \bibfield  {author} {\bibinfo {author} {\bibfnamefont {B.}~\bibnamefont
  {Gertjerenken}}, \bibinfo {author} {\bibfnamefont {T.~P.}\ \bibnamefont
  {Billam}}, \bibinfo {author} {\bibfnamefont {C.~L.}\ \bibnamefont
  {Blackley}}, \bibinfo {author} {\bibfnamefont {C.~R.}\ \bibnamefont
  {Le~Sueur}}, \bibinfo {author} {\bibfnamefont {L.}~\bibnamefont
  {Khaykovich}}, \bibinfo {author} {\bibfnamefont {S.~L.}\ \bibnamefont
  {Cornish}}, \ and\ \bibinfo {author} {\bibfnamefont {C.}~\bibnamefont
  {Weiss}},\ }\bibfield  {title} {\enquote {\bibinfo {title} {Generating
  mesoscopic {Bell} states via collisions of distinguishable quantum bright
  solitons},}\ }\href {\doibase 10.1103/PhysRevLett.111.100406} {\bibfield
  {journal} {\bibinfo  {journal} {Phys. Rev. Lett.}\ }\textbf {\bibinfo
  {volume} {111}},\ \bibinfo {pages} {100406} (\bibinfo {year}
  {2013})}\BibitemShut {NoStop}%
\bibitem [{\citenamefont {Barbiero}\ \emph {et~al.}(2014)\citenamefont
  {Barbiero}, \citenamefont {Malomed},\ and\ \citenamefont
  {Salasnich}}]{BarbieroEtAl2014}%
  \BibitemOpen
  \bibfield  {author} {\bibinfo {author} {\bibfnamefont {L.}~\bibnamefont
  {Barbiero}}, \bibinfo {author} {\bibfnamefont {B.~A.}\ \bibnamefont
  {Malomed}}, \ and\ \bibinfo {author} {\bibfnamefont {L.}~\bibnamefont
  {Salasnich}},\ }\bibfield  {title} {\enquote {\bibinfo {title} {Quantum
  bright solitons in the {Bose-Hubbard} model with site-dependent repulsive
  interactions},}\ }\href {\doibase 10.1103/PhysRevA.90.063611} {\bibfield
  {journal} {\bibinfo  {journal} {Phys. Rev. A}\ }\textbf {\bibinfo {volume}
  {90}},\ \bibinfo {pages} {063611} (\bibinfo {year} {2014})}\BibitemShut
  {NoStop}%
\bibitem [{\citenamefont {Delande}\ and\ \citenamefont
  {Sacha}(2014)}]{DelandeSacha2014}%
  \BibitemOpen
  \bibfield  {author} {\bibinfo {author} {\bibfnamefont {D.}~\bibnamefont
  {Delande}}\ and\ \bibinfo {author} {\bibfnamefont {K.}~\bibnamefont
  {Sacha}},\ }\bibfield  {title} {\enquote {\bibinfo {title} {Many-body
  matter-wave dark soliton},}\ }\href {\doibase 10.1103/PhysRevLett.112.040402}
  {\bibfield  {journal} {\bibinfo  {journal} {Phys. Rev. Lett.}\ }\textbf
  {\bibinfo {volume} {112}},\ \bibinfo {pages} {040402} (\bibinfo {year}
  {2014})}\BibitemShut {NoStop}%
\bibitem [{\citenamefont {Kr\"onke}\ and\ \citenamefont
  {Schmelcher}(2015)}]{KronkeSchmelcher2014}%
  \BibitemOpen
  \bibfield  {author} {\bibinfo {author} {\bibfnamefont {S.}~\bibnamefont
  {Kr\"onke}}\ and\ \bibinfo {author} {\bibfnamefont {P.}~\bibnamefont
  {Schmelcher}},\ }\bibfield  {title} {\enquote {\bibinfo {title} {Many-body
  processes in black and gray matter-wave solitons},}\ }\href {\doibase
  10.1103/PhysRevA.91.053614} {\bibfield  {journal} {\bibinfo  {journal} {Phys.
  Rev. A}\ }\textbf {\bibinfo {volume} {91}},\ \bibinfo {pages} {053614}
  (\bibinfo {year} {2015})}\BibitemShut {NoStop}%
\bibitem [{\citenamefont {Gertjerenken}\ and\ \citenamefont
  {Kevrekidis}(2015)}]{GertjerenkenKevrekidis2015}%
  \BibitemOpen
  \bibfield  {author} {\bibinfo {author} {\bibfnamefont {B.}~\bibnamefont
  {Gertjerenken}}\ and\ \bibinfo {author} {\bibfnamefont {P.G.}\ \bibnamefont
  {Kevrekidis}},\ }\bibfield  {title} {\enquote {\bibinfo {title} {Effects of
  interactions on the generalized {Hong-Ou-Mandel} effect},}\ }\href {\doibase
  http://dx.doi.org/10.1016/j.physleta.2015.04.001} {\bibfield  {journal}
  {\bibinfo  {journal} {Phys. Lett. A}\ }\textbf {\bibinfo {volume} {379}},\
  \bibinfo {pages} {1737} (\bibinfo {year} {2015})}\BibitemShut {NoStop}%
\bibitem [{\citenamefont {Steinigeweg}\ \emph {et~al.}(2007)\citenamefont
  {Steinigeweg}, \citenamefont {Breuer},\ and\ \citenamefont
  {Gemmer}}]{SteinigewegEtAl2007}%
  \BibitemOpen
  \bibfield  {author} {\bibinfo {author} {\bibfnamefont {R.}~\bibnamefont
  {Steinigeweg}}, \bibinfo {author} {\bibfnamefont {H.-P.}\ \bibnamefont
  {Breuer}}, \ and\ \bibinfo {author} {\bibfnamefont {J.}~\bibnamefont
  {Gemmer}},\ }\bibfield  {title} {\enquote {\bibinfo {title} {Transition from
  diffusive to ballistic dynamics for a class of finite quantum models},}\
  }\href {\doibase 10.1103/PhysRevLett.99.150601} {\bibfield  {journal}
  {\bibinfo  {journal} {Phys.\ Rev.\ Lett.}\ }\textbf {\bibinfo {volume}
  {99}},\ \bibinfo {pages} {150601} (\bibinfo {year} {2007})}\BibitemShut
  {NoStop}%
\bibitem [{\citenamefont {{Grabert}}\ \emph {et~al.}(1988)\citenamefont
  {{Grabert}}, \citenamefont {{Schramm}},\ and\ \citenamefont
  {{Ingold}}}]{GrabertEtAl1988}%
  \BibitemOpen
  \bibfield  {author} {\bibinfo {author} {\bibfnamefont {H.}~\bibnamefont
  {{Grabert}}}, \bibinfo {author} {\bibfnamefont {P.}~\bibnamefont
  {{Schramm}}}, \ and\ \bibinfo {author} {\bibfnamefont {G.-L.}\ \bibnamefont
  {{Ingold}}},\ }\bibfield  {title} {\enquote {\bibinfo {title} {{Quantum
  {Brownian} motion: The functional integral approach}},}\ }\href {\doibase
  10.1016/0370-1573(88)90023-3} {\bibfield  {journal} {\bibinfo  {journal}
  {Phys. Rep.}\ }\textbf {\bibinfo {volume} {168}},\ \bibinfo {pages}
  {115--207} (\bibinfo {year} {1988})}\BibitemShut {NoStop}%
\bibitem [{\citenamefont {{Jung}}\ and\ \citenamefont
  {{H{\"a}nggi}}(1991)}]{JungHanggi1991}%
  \BibitemOpen
  \bibfield  {author} {\bibinfo {author} {\bibfnamefont {P.}~\bibnamefont
  {{Jung}}}\ and\ \bibinfo {author} {\bibfnamefont {P.}~\bibnamefont
  {{H{\"a}nggi}}},\ }\bibfield  {title} {\enquote {\bibinfo {title}
  {{Amplification of small signals via stochastic resonance}},}\ }\href
  {\doibase 10.1103/PhysRevA.44.8032} {\bibfield  {journal} {\bibinfo
  {journal} {\pra}\ }\textbf {\bibinfo {volume} {44}},\ \bibinfo {pages} {8032}
  (\bibinfo {year} {1991})}\BibitemShut {NoStop}%
\bibitem [{\citenamefont {Luki\ifmmode~\acute{c}\else \'{c}\fi{}}\ \emph
  {et~al.}(2005)\citenamefont {Luki\ifmmode~\acute{c}\else \'{c}\fi{}},
  \citenamefont {Jeney}, \citenamefont {Tischer}, \citenamefont {Kulik},
  \citenamefont {Forr\'o},\ and\ \citenamefont {Florin}}]{LukicEtAl2005}%
  \BibitemOpen
  \bibfield  {author} {\bibinfo {author} {\bibfnamefont {B.}~\bibnamefont
  {Luki\ifmmode~\acute{c}\else \'{c}\fi{}}}, \bibinfo {author} {\bibfnamefont
  {S.}~\bibnamefont {Jeney}}, \bibinfo {author} {\bibfnamefont
  {C.}~\bibnamefont {Tischer}}, \bibinfo {author} {\bibfnamefont {A.~J.}\
  \bibnamefont {Kulik}}, \bibinfo {author} {\bibfnamefont {L.}~\bibnamefont
  {Forr\'o}}, \ and\ \bibinfo {author} {\bibfnamefont {E.-L.}\ \bibnamefont
  {Florin}},\ }\bibfield  {title} {\enquote {\bibinfo {title} {Direct
  observation of nondiffusive motion of a {Brownian} particle},}\ }\href
  {\doibase 10.1103/PhysRevLett.95.160601} {\bibfield  {journal} {\bibinfo
  {journal} {Phys. Rev. Lett.}\ }\textbf {\bibinfo {volume} {95}},\ \bibinfo
  {pages} {160601} (\bibinfo {year} {2005})}\BibitemShut {NoStop}%
\bibitem [{\citenamefont {K\"oppl}\ \emph {et~al.}(2006)\citenamefont
  {K\"oppl}, \citenamefont {Henseler}, \citenamefont {Erbe}, \citenamefont
  {Nielaba},\ and\ \citenamefont {Leiderer}}]{KoepplEtAl2006}%
  \BibitemOpen
  \bibfield  {author} {\bibinfo {author} {\bibfnamefont {M.}~\bibnamefont
  {K\"oppl}}, \bibinfo {author} {\bibfnamefont {P.}~\bibnamefont {Henseler}},
  \bibinfo {author} {\bibfnamefont {A.}~\bibnamefont {Erbe}}, \bibinfo {author}
  {\bibfnamefont {P.}~\bibnamefont {Nielaba}}, \ and\ \bibinfo {author}
  {\bibfnamefont {P.}~\bibnamefont {Leiderer}},\ }\bibfield  {title} {\enquote
  {\bibinfo {title} {Layer reduction in driven 2d-colloidal systems through
  microchannels},}\ }\href {\doibase 10.1103/PhysRevLett.97.208302} {\bibfield
  {journal} {\bibinfo  {journal} {Phys. Rev. Lett.}\ }\textbf {\bibinfo
  {volume} {97}},\ \bibinfo {pages} {208302} (\bibinfo {year}
  {2006})}\BibitemShut {NoStop}%
\bibitem [{\citenamefont {H\"anggi}\ and\ \citenamefont
  {Marchesoni}(2009)}]{HanggiEtAl2009}%
  \BibitemOpen
  \bibfield  {author} {\bibinfo {author} {\bibfnamefont {P.}~\bibnamefont
  {H\"anggi}}\ and\ \bibinfo {author} {\bibfnamefont {F.}~\bibnamefont
  {Marchesoni}},\ }\bibfield  {title} {\enquote {\bibinfo {title} {Artificial
  brownian motors: Controlling transport on the nanoscale},}\ }\href {\doibase
  10.1103/RevModPhys.81.387} {\bibfield  {journal} {\bibinfo  {journal} {Rev.
  Mod. Phys.}\ }\textbf {\bibinfo {volume} {81}},\ \bibinfo {pages} {387}
  (\bibinfo {year} {2009})}\BibitemShut {NoStop}%
\bibitem [{\citenamefont {Dierl}\ \emph {et~al.}(2014)\citenamefont {Dierl},
  \citenamefont {Dieterich}, \citenamefont {Einax},\ and\ \citenamefont
  {Maass}}]{DierlEtAl2014}%
  \BibitemOpen
  \bibfield  {author} {\bibinfo {author} {\bibfnamefont {M.}~\bibnamefont
  {Dierl}}, \bibinfo {author} {\bibfnamefont {W.}~\bibnamefont {Dieterich}},
  \bibinfo {author} {\bibfnamefont {M.}~\bibnamefont {Einax}}, \ and\ \bibinfo
  {author} {\bibfnamefont {P.}~\bibnamefont {Maass}},\ }\bibfield  {title}
  {\enquote {\bibinfo {title} {Phase transitions in {Brownian} pumps},}\ }\href
  {\doibase 10.1103/PhysRevLett.112.150601} {\bibfield  {journal} {\bibinfo
  {journal} {Phys. Rev. Lett.}\ }\textbf {\bibinfo {volume} {112}},\ \bibinfo
  {pages} {150601} (\bibinfo {year} {2014})}\BibitemShut {NoStop}%
\bibitem [{\citenamefont {{Turiv}}\ \emph {et~al.}(2015)\citenamefont
  {{Turiv}}, \citenamefont {{Brodin}},\ and\ \citenamefont
  {{Nazarenko}}}]{TurivEtAl2015}%
  \BibitemOpen
  \bibfield  {author} {\bibinfo {author} {\bibfnamefont {T.}~\bibnamefont
  {{Turiv}}}, \bibinfo {author} {\bibfnamefont {A.}~\bibnamefont {{Brodin}}}, \
  and\ \bibinfo {author} {\bibfnamefont {V.}~\bibnamefont {{Nazarenko}}},\
  }\bibfield  {title} {\enquote {\bibinfo {title} {{Anomalous {Brownian} motion
  of colloidal particle in a nematic environment: effect of the director
  fluctuations}},}\ }\href@noop {} {\bibfield  {journal} {\bibinfo  {journal}
  {Condens. Matter Phys.}\ }\textbf {\bibinfo {volume} {18}},\ \bibinfo {pages}
  {23001} (\bibinfo {year} {2015})}\BibitemShut {NoStop}%
\bibitem [{\citenamefont {Fisher}\ and\ \citenamefont
  {Zwerger}(1985)}]{FisherZwerger1985}%
  \BibitemOpen
  \bibfield  {author} {\bibinfo {author} {\bibfnamefont {M.~P.~A.}\
  \bibnamefont {Fisher}}\ and\ \bibinfo {author} {\bibfnamefont
  {W.}~\bibnamefont {Zwerger}},\ }\bibfield  {title} {\enquote {\bibinfo
  {title} {Quantum {Brownian} motion in a periodic potential},}\ }\href
  {\doibase 10.1103/PhysRevB.32.6190} {\bibfield  {journal} {\bibinfo
  {journal} {Phys. Rev. B}\ }\textbf {\bibinfo {volume} {32}},\ \bibinfo
  {pages} {6190} (\bibinfo {year} {1985})}\BibitemShut {NoStop}%
\bibitem [{\citenamefont {Metzler}\ and\ \citenamefont
  {Klafter}(2000)}]{Metzler2000}%
  \BibitemOpen
  \bibfield  {author} {\bibinfo {author} {\bibfnamefont {R.}~\bibnamefont
  {Metzler}}\ and\ \bibinfo {author} {\bibfnamefont {J.}~\bibnamefont
  {Klafter}},\ }\bibfield  {title} {\enquote {\bibinfo {title} {The random
  walk's guide to anomalous diffusion: a fractional dynamics approach},}\
  }\href {\doibase http://dx.doi.org/10.1016/S0370-1573(00)00070-3} {\bibfield
  {journal} {\bibinfo  {journal} {Phys. Rep.}\ }\textbf {\bibinfo {volume}
  {339}},\ \bibinfo {pages} {1} (\bibinfo {year} {2000})}\BibitemShut {NoStop}%
\bibitem [{\citenamefont {D\"ur}\ \emph {et~al.}(2002)\citenamefont {D\"ur},
  \citenamefont {Raussendorf}, \citenamefont {Kendon},\ and\ \citenamefont
  {Briegel}}]{DurEtAl2002}%
  \BibitemOpen
  \bibfield  {author} {\bibinfo {author} {\bibfnamefont {W.}~\bibnamefont
  {D\"ur}}, \bibinfo {author} {\bibfnamefont {R.}~\bibnamefont {Raussendorf}},
  \bibinfo {author} {\bibfnamefont {V.~M.}\ \bibnamefont {Kendon}}, \ and\
  \bibinfo {author} {\bibfnamefont {H.-J.}\ \bibnamefont {Briegel}},\
  }\bibfield  {title} {\enquote {\bibinfo {title} {Quantum walks in optical
  lattices},}\ }\href {\doibase 10.1103/PhysRevA.66.052319} {\bibfield
  {journal} {\bibinfo  {journal} {Phys. Rev. A}\ }\textbf {\bibinfo {volume}
  {66}},\ \bibinfo {pages} {052319} (\bibinfo {year} {2002})}\BibitemShut
  {NoStop}%
\bibitem [{\citenamefont {Karski}\ \emph {et~al.}(2009)\citenamefont {Karski},
  \citenamefont {F{\"o}rster}, \citenamefont {Choi}, \citenamefont {Steffen},
  \citenamefont {Alt}, \citenamefont {Meschede},\ and\ \citenamefont
  {Widera}}]{KarskiEtAl2009}%
  \BibitemOpen
  \bibfield  {author} {\bibinfo {author} {\bibfnamefont {M.}~\bibnamefont
  {Karski}}, \bibinfo {author} {\bibfnamefont {L.}~\bibnamefont {F{\"o}rster}},
  \bibinfo {author} {\bibfnamefont {J.-M.}\ \bibnamefont {Choi}}, \bibinfo
  {author} {\bibfnamefont {A.}~\bibnamefont {Steffen}}, \bibinfo {author}
  {\bibfnamefont {W.}~\bibnamefont {Alt}}, \bibinfo {author} {\bibfnamefont
  {D.}~\bibnamefont {Meschede}}, \ and\ \bibinfo {author} {\bibfnamefont
  {A.}~\bibnamefont {Widera}},\ }\bibfield  {title} {\enquote {\bibinfo {title}
  {Quantum walk in position space with single optically trapped atoms},}\
  }\href {\doibase 10.1126/science.1174436} {\bibfield  {journal} {\bibinfo
  {journal} {Science}\ }\textbf {\bibinfo {volume} {325}},\ \bibinfo {pages}
  {174} (\bibinfo {year} {2009})}\BibitemShut {NoStop}%
\bibitem [{\citenamefont {Agliari}\ \emph {et~al.}(2010)\citenamefont
  {Agliari}, \citenamefont {Blumen},\ and\ \citenamefont
  {M\"ulken}}]{AgliariEtAl2010}%
  \BibitemOpen
  \bibfield  {author} {\bibinfo {author} {\bibfnamefont {E.}~\bibnamefont
  {Agliari}}, \bibinfo {author} {\bibfnamefont {A.}~\bibnamefont {Blumen}}, \
  and\ \bibinfo {author} {\bibfnamefont {O.}~\bibnamefont {M\"ulken}},\
  }\bibfield  {title} {\enquote {\bibinfo {title} {Quantum-walk approach to
  searching on fractal structures},}\ }\href {\doibase
  10.1103/PhysRevA.82.012305} {\bibfield  {journal} {\bibinfo  {journal} {Phys.
  Rev. A}\ }\textbf {\bibinfo {volume} {82}},\ \bibinfo {pages} {012305}
  (\bibinfo {year} {2010})}\BibitemShut {NoStop}%
\bibitem [{\citenamefont {Olshanii}(1998)}]{Olshanii1998}%
  \BibitemOpen
  \bibfield  {author} {\bibinfo {author} {\bibfnamefont {M.}~\bibnamefont
  {Olshanii}},\ }\bibfield  {title} {\enquote {\bibinfo {title} {Atomic
  scattering in the presence of an external confinement and a gas of
  impenetrable bosons},}\ }\href {\doibase 10.1103/PhysRevLett.81.938}
  {\bibfield  {journal} {\bibinfo  {journal} {Phys. Rev. Lett.}\ }\textbf
  {\bibinfo {volume} {81}},\ \bibinfo {pages} {938} (\bibinfo {year}
  {1998})}\BibitemShut {NoStop}%
\bibitem [{\citenamefont {Calogero}\ and\ \citenamefont
  {Degasperis}(1975)}]{CalogeroDegasperis1975}%
  \BibitemOpen
  \bibfield  {author} {\bibinfo {author} {\bibfnamefont {F.}~\bibnamefont
  {Calogero}}\ and\ \bibinfo {author} {\bibfnamefont {A.}~\bibnamefont
  {Degasperis}},\ }\bibfield  {title} {\enquote {\bibinfo {title} {{Comparison
  between the exact and Hartree solutions of a one-dimensional many-body
  problem}},}\ }\href {\doibase 10.1103/PhysRevA.11.265} {\bibfield  {journal}
  {\bibinfo  {journal} {Phys. Rev. A}\ }\textbf {\bibinfo {volume} {11}},\
  \bibinfo {pages} {265} (\bibinfo {year} {1975})}\BibitemShut {NoStop}%
\bibitem [{\citenamefont {Castin}\ and\ \citenamefont
  {Herzog}(2001)}]{CastinHerzog2001}%
  \BibitemOpen
  \bibfield  {author} {\bibinfo {author} {\bibfnamefont {Y.}~\bibnamefont
  {Castin}}\ and\ \bibinfo {author} {\bibfnamefont {C.}~\bibnamefont
  {Herzog}},\ }\bibfield  {title} {\enquote {\bibinfo {title} {{Bose-Einstein}
  condensates in symmetry breaking states},}\ }\href {\doibase
  10.1016/S1296-2147(01)01183-0} {\bibfield  {journal} {\bibinfo  {journal} {C.
  R. Acad. Sci. Paris, Ser. IV}\ }\textbf {\bibinfo {volume} {2}},\ \bibinfo
  {pages} {419} (\bibinfo {year} {2001})},\ \Eprint
  {http://arxiv.org/abs/arXiv:cond-mat/0012040} {arXiv:cond-mat/0012040}
  \BibitemShut {NoStop}%
\bibitem [{\citenamefont {Lieb}\ and\ \citenamefont
  {Liniger}(1963)}]{LiebLiniger1963}%
  \BibitemOpen
  \bibfield  {author} {\bibinfo {author} {\bibfnamefont {E.~H.}\ \bibnamefont
  {Lieb}}\ and\ \bibinfo {author} {\bibfnamefont {W.}~\bibnamefont {Liniger}},\
  }\bibfield  {title} {\enquote {\bibinfo {title} {{Exact Analysis of an
  Interacting Bose Gas. I. The General Solution and the Ground State}},}\
  }\href {\doibase 10.1103/PhysRev.130.1605} {\bibfield  {journal} {\bibinfo
  {journal} {Phys. Rev.}\ }\textbf {\bibinfo {volume} {130}},\ \bibinfo {pages}
  {1605} (\bibinfo {year} {1963})}\BibitemShut {NoStop}%
\bibitem [{\citenamefont {Castin}(2009)}]{Castin2009}%
  \BibitemOpen
  \bibfield  {author} {\bibinfo {author} {\bibfnamefont {Y.}~\bibnamefont
  {Castin}},\ }\bibfield  {title} {\enquote {\bibinfo {title} {Internal
  structure of a quantum soliton and classical excitations due to trap
  opening},}\ }\href {\doibase 10.1140/epjb/e2008-00407-3} {\bibfield
  {journal} {\bibinfo  {journal} {Eur. Phys. J. B}\ }\textbf {\bibinfo {volume}
  {68}},\ \bibinfo {pages} {317} (\bibinfo {year} {2009})}\BibitemShut
  {NoStop}%
\bibitem [{\citenamefont {Holdaway}\ \emph {et~al.}(2012)\citenamefont
  {Holdaway}, \citenamefont {Weiss},\ and\ \citenamefont
  {Gardiner}}]{HoldawayEtAl2012}%
  \BibitemOpen
  \bibfield  {author} {\bibinfo {author} {\bibfnamefont {D.~I.~H.}\
  \bibnamefont {Holdaway}}, \bibinfo {author} {\bibfnamefont {C.}~\bibnamefont
  {Weiss}}, \ and\ \bibinfo {author} {\bibfnamefont {S.~A.}\ \bibnamefont
  {Gardiner}},\ }\bibfield  {title} {\enquote {\bibinfo {title} {Quantum theory
  of bright matter-wave solitons in harmonic confinement},}\ }\href {\doibase
  10.1103/PhysRevA.85.053618} {\bibfield  {journal} {\bibinfo  {journal} {Phys.
  Rev. A}\ }\textbf {\bibinfo {volume} {85}},\ \bibinfo {pages} {053618}
  (\bibinfo {year} {2012})}\BibitemShut {NoStop}%
\bibitem [{\citenamefont {Bonitz}\ \emph {et~al.}(2007)\citenamefont {Bonitz},
  \citenamefont {Balzer},\ and\ \citenamefont {van Leeuwen}}]{BonitzEtAl2007}%
  \BibitemOpen
  \bibfield  {author} {\bibinfo {author} {\bibfnamefont {M.}~\bibnamefont
  {Bonitz}}, \bibinfo {author} {\bibfnamefont {K.}~\bibnamefont {Balzer}}, \
  and\ \bibinfo {author} {\bibfnamefont {R.}~\bibnamefont {van Leeuwen}},\
  }\bibfield  {title} {\enquote {\bibinfo {title} {Invariance of the kohn
  center-of-mass mode in a conserving theory},}\ }\href {\doibase
  10.1103/PhysRevB.76.045341} {\bibfield  {journal} {\bibinfo  {journal} {Phys.
  Rev. B}\ }\textbf {\bibinfo {volume} {76}},\ \bibinfo {pages} {045341}
  (\bibinfo {year} {2007})}\BibitemShut {NoStop}%
\bibitem [{\citenamefont {{Fl\"ugge}}(1990)}]{Fluegge1990}%
  \BibitemOpen
  \bibfield  {author} {\bibinfo {author} {\bibfnamefont {S.}~\bibnamefont
  {{Fl\"ugge}}},\ }\href@noop {} {\emph {\bibinfo {title} {Rechenmethoden der
  Quantentheorie}}}\ (\bibinfo  {publisher} {Springer},\ \bibinfo {address}
  {Berlin},\ \bibinfo {year} {1990})\BibitemShut {NoStop}%
\bibitem [{\citenamefont {Sinatra}\ \emph {et~al.}(2002)\citenamefont
  {Sinatra}, \citenamefont {Lobo},\ and\ \citenamefont
  {Castin}}]{SinatraEtAl2002}%
  \BibitemOpen
  \bibfield  {author} {\bibinfo {author} {\bibfnamefont {A.}~\bibnamefont
  {Sinatra}}, \bibinfo {author} {\bibfnamefont {C.}~\bibnamefont {Lobo}}, \
  and\ \bibinfo {author} {\bibfnamefont {Y.}~\bibnamefont {Castin}},\
  }\bibfield  {title} {\enquote {\bibinfo {title} {The truncated {Wigner}
  method for {Bose-condensed} gases: limits of validity and applications},}\
  }\href {\doibase 10.1088/0953-4075/35/17/301} {\bibfield  {journal} {\bibinfo
   {journal} {J. Phys. B}\ }\textbf {\bibinfo {volume} {35}},\ \bibinfo {pages}
  {3599} (\bibinfo {year} {2002})}\BibitemShut {NoStop}%
\bibitem [{\citenamefont {Bienias}\ \emph {et~al.}(2011)\citenamefont
  {Bienias}, \citenamefont {Pawlowski}, \citenamefont {Gajda},\ and\
  \citenamefont {Rzazewski}}]{BieniasEtAl2011}%
  \BibitemOpen
  \bibfield  {author} {\bibinfo {author} {\bibfnamefont {P.}~\bibnamefont
  {Bienias}}, \bibinfo {author} {\bibfnamefont {K.}~\bibnamefont {Pawlowski}},
  \bibinfo {author} {\bibfnamefont {M.}~\bibnamefont {Gajda}}, \ and\ \bibinfo
  {author} {\bibfnamefont {K.}~\bibnamefont {Rzazewski}},\ }\bibfield  {title}
  {\enquote {\bibinfo {title} {Statistical properties of one-dimensional
  attractive {Bose} gas},}\ }\href {\doibase 10.1209/0295-5075/96/10011}
  {\bibfield  {journal} {\bibinfo  {journal} {EPL (Europhys. Lett.)}\ }\textbf
  {\bibinfo {volume} {96}},\ \bibinfo {pages} {10011} (\bibinfo {year}
  {2011})}\BibitemShut {NoStop}%
\bibitem [{\citenamefont {Davis}(1993)}]{Davis1993}%
  \BibitemOpen
  \bibfield  {author} {\bibinfo {author} {\bibfnamefont {M.~H.~A.}\
  \bibnamefont {Davis}},\ }\href@noop {} {\emph {\bibinfo {title} {Markov
  models and optimization}}}\ (\bibinfo  {publisher} {Chapman {\&} Hall},\
  \bibinfo {address} {London},\ \bibinfo {year} {1993})\BibitemShut {NoStop}%
\bibitem [{\citenamefont {Dalibard}\ \emph {et~al.}(1992)\citenamefont
  {Dalibard}, \citenamefont {Castin},\ and\ \citenamefont
  {M{\o}lmer}}]{DalibardEtAl1992}%
  \BibitemOpen
  \bibfield  {author} {\bibinfo {author} {\bibfnamefont {J.}~\bibnamefont
  {Dalibard}}, \bibinfo {author} {\bibfnamefont {Y.}~\bibnamefont {Castin}}, \
  and\ \bibinfo {author} {\bibfnamefont {K.}~\bibnamefont {M{\o}lmer}},\
  }\bibfield  {title} {\enquote {\bibinfo {title} {Wave-function approach to
  dissipative processes in quantum optics},}\ }\href {\doibase
  10.1103/PhysRevLett.68.580} {\bibfield  {journal} {\bibinfo  {journal}
  {Phys.\ Rev.\ Lett.}\ }\textbf {\bibinfo {volume} {68}},\ \bibinfo {pages}
  {580} (\bibinfo {year} {1992})}\BibitemShut {NoStop}%
\bibitem [{\citenamefont {Dum}\ \emph {et~al.}(1992)\citenamefont {Dum},
  \citenamefont {Zoller},\ and\ \citenamefont {Ritsch}}]{DumEtAl1992}%
  \BibitemOpen
  \bibfield  {author} {\bibinfo {author} {\bibfnamefont {R.}~\bibnamefont
  {Dum}}, \bibinfo {author} {\bibfnamefont {P.}~\bibnamefont {Zoller}}, \ and\
  \bibinfo {author} {\bibfnamefont {H.}~\bibnamefont {Ritsch}},\ }\bibfield
  {title} {\enquote {\bibinfo {title} {{Monte Carlo} simulation of the atomic
  master equation for spontaneous emission},}\ }\href {\doibase
  10.1103/PhysRevA.45.4879} {\bibfield  {journal} {\bibinfo  {journal} {Phys.
  Rev. A}\ }\textbf {\bibinfo {volume} {45}},\ \bibinfo {pages} {4879}
  (\bibinfo {year} {1992})}\BibitemShut {NoStop}%
\bibitem [{\citenamefont {Breuer}\ and\ \citenamefont
  {Petruccione}(2006)}]{Breuer2006}%
  \BibitemOpen
  \bibfield  {author} {\bibinfo {author} {\bibfnamefont {H.-P.}\ \bibnamefont
  {Breuer}}\ and\ \bibinfo {author} {\bibfnamefont {F.}~\bibnamefont
  {Petruccione}},\ }\href@noop {} {\emph {\bibinfo {title} {The Theory of Open
  Quantum Systems}}}\ (\bibinfo  {publisher} {Clarendon Press},\ \bibinfo
  {address} {Oxford},\ \bibinfo {year} {2006})\BibitemShut {NoStop}%
\bibitem [{\citenamefont {Kr{\"o}nke}\ \emph {et~al.}(2015)\citenamefont
  {Kr{\"o}nke}, \citenamefont {Kn{\"o}rzer},\ and\ \citenamefont
  {Schmelcher}}]{KronkeEtAl2015}%
  \BibitemOpen
  \bibfield  {author} {\bibinfo {author} {\bibfnamefont {S.}~\bibnamefont
  {Kr{\"o}nke}}, \bibinfo {author} {\bibfnamefont {J.}~\bibnamefont
  {Kn{\"o}rzer}}, \ and\ \bibinfo {author} {\bibfnamefont {P.}~\bibnamefont
  {Schmelcher}},\ }\bibfield  {title} {\enquote {\bibinfo {title} {Correlated
  quantum dynamics of a single atom collisionally coupled to an ultracold
  finite bosonic ensemble},}\ }\href {\doibase 10.1088/1367-2630/17/5/053001}
  {\bibfield  {journal} {\bibinfo  {journal} {New J. Phys.}\ }\textbf {\bibinfo
  {volume} {17}},\ \bibinfo {pages} {053001} (\bibinfo {year}
  {2015})}\BibitemShut {NoStop}%
\bibitem [{\citenamefont {{Grimm}}\ \emph {et~al.}(2000)\citenamefont
  {{Grimm}}, \citenamefont {{Weidem{\"u}ller}},\ and\ \citenamefont
  {{Ovchinnikov}}}]{GrimmEtAl2000}%
  \BibitemOpen
  \bibfield  {author} {\bibinfo {author} {\bibfnamefont {R.}~\bibnamefont
  {{Grimm}}}, \bibinfo {author} {\bibfnamefont {M.}~\bibnamefont
  {{Weidem{\"u}ller}}}, \ and\ \bibinfo {author} {\bibfnamefont {Y.~B.}\
  \bibnamefont {{Ovchinnikov}}},\ }\bibfield  {title} {\enquote {\bibinfo
  {title} {{Optical Dipole Traps for Neutral Atoms}},}\ }\href {\doibase
  10.1016/S1049-250X(08)60186-X} {\bibfield  {journal} {\bibinfo  {journal}
  {Adv. At. Mol. Opt. Phys.}\ }\textbf {\bibinfo {volume} {42}},\ \bibinfo
  {pages} {95} (\bibinfo {year} {2000})},\ \Eprint
  {http://arxiv.org/abs/physics/9902072} {physics/9902072} \BibitemShut
  {NoStop}%
\bibitem [{map()}]{maple}%
  \BibitemOpen
  \href@noop {} {}\bibinfo {note} {{Computer algebra programme \textsc{MAPLE},
  \url{http://www.maplesoft.com/}}}\BibitemShut {NoStop}%
\bibitem [{\citenamefont {Shotan}\ \emph {et~al.}(2014)\citenamefont {Shotan},
  \citenamefont {Machtey}, \citenamefont {Kokkelmans},\ and\ \citenamefont
  {Khaykovich}}]{ShotanEtAl2014}%
  \BibitemOpen
  \bibfield  {author} {\bibinfo {author} {\bibfnamefont {Z.}~\bibnamefont
  {Shotan}}, \bibinfo {author} {\bibfnamefont {O.}~\bibnamefont {Machtey}},
  \bibinfo {author} {\bibfnamefont {S.}~\bibnamefont {Kokkelmans}}, \ and\
  \bibinfo {author} {\bibfnamefont {L.}~\bibnamefont {Khaykovich}},\ }\bibfield
   {title} {\enquote {\bibinfo {title} {Three-body recombination at vanishing
  scattering lengths in an ultracold bose gas},}\ }\href {\doibase
  10.1103/PhysRevLett.113.053202} {\bibfield  {journal} {\bibinfo  {journal}
  {Phys. Rev. Lett.}\ }\textbf {\bibinfo {volume} {113}},\ \bibinfo {pages}
  {053202} (\bibinfo {year} {2014})}\BibitemShut {NoStop}%
\bibitem [{\citenamefont {Roberts}\ \emph {et~al.}(2000)\citenamefont
  {Roberts}, \citenamefont {Claussen}, \citenamefont {Cornish},\ and\
  \citenamefont {Wieman}}]{RobertsEtAl2000}%
  \BibitemOpen
  \bibfield  {author} {\bibinfo {author} {\bibfnamefont {J.~L.}\ \bibnamefont
  {Roberts}}, \bibinfo {author} {\bibfnamefont {N.~R.}\ \bibnamefont
  {Claussen}}, \bibinfo {author} {\bibfnamefont {S.~L.}\ \bibnamefont
  {Cornish}}, \ and\ \bibinfo {author} {\bibfnamefont {C.~E.}\ \bibnamefont
  {Wieman}},\ }\bibfield  {title} {\enquote {\bibinfo {title} {Magnetic field
  dependence of ultracold inelastic collisions near a {Feshbach} resonance},}\
  }\href {\doibase 10.1103/PhysRevLett.85.728} {\bibfield  {journal} {\bibinfo
  {journal} {Phys. Rev. Lett.}\ }\textbf {\bibinfo {volume} {85}},\ \bibinfo
  {pages} {728} (\bibinfo {year} {2000})}\BibitemShut {NoStop}%
\bibitem [{\citenamefont {Weiss}\ \emph
  {et~al.}(2015{\natexlab{b}})\citenamefont {Weiss}, \citenamefont {Cornish},
  \citenamefont {Gardiner},\ and\ \citenamefont {Breuer}}]{WeissEtAl2015bData}%
  \BibitemOpen
  \bibfield  {author} {\bibinfo {author} {\bibfnamefont {C.}~\bibnamefont
  {Weiss}}, \bibinfo {author} {\bibfnamefont {S.~L.}\ \bibnamefont {Cornish}},
  \bibinfo {author} {\bibfnamefont {S.~A.}\ \bibnamefont {Gardiner}}, \ and\
  \bibinfo {author} {\bibfnamefont {H.-P.}\ \bibnamefont {Breuer}},\
  }\href@noop {} {}\bibinfo {howpublished}
  {\url{https://collections.durham.ac.uk/files/44558d350} and
  \url{http://dx.doi.org/10.15128/44558d350}} (\bibinfo {year}
  {2015}{\natexlab{b}}),\ \bibinfo {note} {{``Superballistic center-of-mass
  motion in one-dimensional attractive Bose gases: Supporting
  Data''}}\BibitemShut {NoStop}%
\bibitem [{\citenamefont {Weiss}\ and\ \citenamefont
  {Castin}(2009)}]{WeissCastin2009}%
  \BibitemOpen
  \bibfield  {author} {\bibinfo {author} {\bibfnamefont {C.}~\bibnamefont
  {Weiss}}\ and\ \bibinfo {author} {\bibfnamefont {Y.}~\bibnamefont {Castin}},\
  }\bibfield  {title} {\enquote {\bibinfo {title} {Creation and detection of a
  mesoscopic gas in a nonlocal quantum superposition},}\ }\href {\doibase
  10.1103/PhysRevLett.102.010403} {\bibfield  {journal} {\bibinfo  {journal}
  {Phys.\ Rev.\ Lett.}\ }\textbf {\bibinfo {volume} {102}},\ \bibinfo {pages}
  {010403} (\bibinfo {year} {2009})}\BibitemShut {NoStop}%
\bibitem [{\citenamefont {Maucher}\ \emph {et~al.}(2012)\citenamefont
  {Maucher}, \citenamefont {Krolikowski},\ and\ \citenamefont
  {Skupin}}]{MaucherEtAl2012}%
  \BibitemOpen
  \bibfield  {author} {\bibinfo {author} {\bibfnamefont {F.}~\bibnamefont
  {Maucher}}, \bibinfo {author} {\bibfnamefont {W.}~\bibnamefont
  {Krolikowski}}, \ and\ \bibinfo {author} {\bibfnamefont {S.}~\bibnamefont
  {Skupin}},\ }\bibfield  {title} {\enquote {\bibinfo {title} {Stability of
  solitary waves in random nonlocal nonlinear media},}\ }\href {\doibase
  10.1103/PhysRevA.85.063803} {\bibfield  {journal} {\bibinfo  {journal} {Phys.
  Rev. A}\ }\textbf {\bibinfo {volume} {85}},\ \bibinfo {pages} {063803}
  (\bibinfo {year} {2012})}\BibitemShut {NoStop}%
\bibitem [{\citenamefont {{Streltsova}}\ and\ \citenamefont
  {{Streltsov}}(2014)}]{StreltsovaStreltsov2014}%
  \BibitemOpen
  \bibfield  {author} {\bibinfo {author} {\bibfnamefont {O.~I.}\ \bibnamefont
  {{Streltsova}}}\ and\ \bibinfo {author} {\bibfnamefont {A.~I.}\ \bibnamefont
  {{Streltsov}}},\ }\bibfield  {title} {\enquote {\bibinfo {title}
  {{Interferometry with correlated matter-waves}},}\ }\href@noop {} {\bibfield
  {journal} {\bibinfo  {journal} {ArXiv e-prints}\ } (\bibinfo {year}
  {2014})},\ \Eprint {http://arxiv.org/abs/1412.4049} {arXiv:1412.4049
  [quant-ph]} \BibitemShut {NoStop}%
\bibitem [{\citenamefont {Martin}(2016)}]{Martin2016}%
  \BibitemOpen
  \bibfield  {author} {\bibinfo {author} {\bibfnamefont {A.~D.}\ \bibnamefont
  {Martin}},\ }\bibfield  {title} {\enquote {\bibinfo {title}
  {Collision-induced frequency shifts in bright matter-wave solitons and
  soliton molecules},}\ }\href {\doibase 10.1103/PhysRevA.93.023631} {\bibfield
   {journal} {\bibinfo  {journal} {Phys. Rev. A}\ }\textbf {\bibinfo {volume}
  {93}},\ \bibinfo {pages} {023631} (\bibinfo {year} {2016})}\BibitemShut
  {NoStop}%
\bibitem [{\citenamefont {Holdaway}\ \emph {et~al.}(2013)\citenamefont
  {Holdaway}, \citenamefont {Weiss},\ and\ \citenamefont
  {Gardiner}}]{HoldawayEtAl2013}%
  \BibitemOpen
  \bibfield  {author} {\bibinfo {author} {\bibfnamefont {D.~I.~H.}\
  \bibnamefont {Holdaway}}, \bibinfo {author} {\bibfnamefont {C.}~\bibnamefont
  {Weiss}}, \ and\ \bibinfo {author} {\bibfnamefont {S.~A.}\ \bibnamefont
  {Gardiner}},\ }\bibfield  {title} {\enquote {\bibinfo {title} {Collision
  dynamics and entanglement generation of two initially independent and
  indistinguishable boson pairs in one-dimensional harmonic confinement},}\
  }\href {\doibase 10.1103/PhysRevA.87.043632} {\bibfield  {journal} {\bibinfo
  {journal} {Phys. Rev. A}\ }\textbf {\bibinfo {volume} {87}},\ \bibinfo
  {pages} {043632} (\bibinfo {year} {2013})}\BibitemShut {NoStop}%
\bibitem [{\citenamefont {{Bowman}}\ \emph {et~al.}(2015)\citenamefont
  {{Bowman}}, \citenamefont {{Ireland}}, \citenamefont {{Bruce}},\ and\
  \citenamefont {{Cassettari}}}]{BowmanEtAl2015}%
  \BibitemOpen
  \bibfield  {author} {\bibinfo {author} {\bibfnamefont {D.}~\bibnamefont
  {{Bowman}}}, \bibinfo {author} {\bibfnamefont {P.}~\bibnamefont {{Ireland}}},
  \bibinfo {author} {\bibfnamefont {G.~D.}\ \bibnamefont {{Bruce}}}, \ and\
  \bibinfo {author} {\bibfnamefont {D.}~\bibnamefont {{Cassettari}}},\
  }\bibfield  {title} {\enquote {\bibinfo {title} {{Multi-wavelength holography
  with a single spatial light modulator for ultracold atom experiments}},}\
  }\href {\doibase 10.1364/OE.23.008365} {\bibfield  {journal} {\bibinfo
  {journal} {Opt. Express}\ }\textbf {\bibinfo {volume} {23}},\ \bibinfo
  {pages} {8365} (\bibinfo {year} {2015})}\BibitemShut {NoStop}%
\bibitem [{\citenamefont {McGuire}(1964)}]{McGuire1964}%
  \BibitemOpen
  \bibfield  {author} {\bibinfo {author} {\bibfnamefont {J.~B.}\ \bibnamefont
  {McGuire}},\ }\bibfield  {title} {\enquote {\bibinfo {title} {{Study of
  Exactly Soluble One-Dimensional N-Body Problems}},}\ }\href {\doibase
  10.1063/1.1704156} {\bibfield  {journal} {\bibinfo  {journal} {J. Math.
  Phys.}\ }\textbf {\bibinfo {volume} {5}},\ \bibinfo {pages} {622} (\bibinfo
  {year} {1964})}\BibitemShut {NoStop}%
\bibitem [{\citenamefont {Obukhov}(1959)}]{Obukhov1959}%
  \BibitemOpen
  \bibfield  {author} {\bibinfo {author} {\bibfnamefont {A.M.}\ \bibnamefont
  {Obukhov}},\ }\bibfield  {title} {\enquote {\bibinfo {title} {Description of
  turbulence in terms of lagrangian variables},}\ \ }(\bibinfo  {publisher}
  {Elsevier},\ \bibinfo {year} {1959})\ p.\ \bibinfo {pages} {113}\BibitemShut
  {NoStop}%
\bibitem [{\citenamefont {Baule}\ and\ \citenamefont
  {Friedrich}(2006)}]{BauleFriedrich2006}%
  \BibitemOpen
  \bibfield  {author} {\bibinfo {author} {\bibfnamefont {A.}~\bibnamefont
  {Baule}}\ and\ \bibinfo {author} {\bibfnamefont {R.}~\bibnamefont
  {Friedrich}},\ }\bibfield  {title} {\enquote {\bibinfo {title} {Investigation
  of a generalized obukhov model for turbulence},}\ }\href {\doibase
  http://dx.doi.org/10.1016/j.physleta.2005.10.017} {\bibfield  {journal}
  {\bibinfo  {journal} {Phys. Lett. A}\ }\textbf {\bibinfo {volume} {350}},\
  \bibinfo {pages} {167} (\bibinfo {year} {2006})}\BibitemShut {NoStop}%
\bibitem [{\citenamefont {Kessler}\ and\ \citenamefont
  {Barkai}(2012)}]{KesslerBarkai2012}%
  \BibitemOpen
  \bibfield  {author} {\bibinfo {author} {\bibfnamefont {David~A.}\
  \bibnamefont {Kessler}}\ and\ \bibinfo {author} {\bibfnamefont {Eli}\
  \bibnamefont {Barkai}},\ }\bibfield  {title} {\enquote {\bibinfo {title}
  {Theory of fractional l\'evy kinetics for cold atoms diffusing in optical
  lattices},}\ }\href {\doibase 10.1103/PhysRevLett.108.230602} {\bibfield
  {journal} {\bibinfo  {journal} {Phys. Rev. Lett.}\ }\textbf {\bibinfo
  {volume} {108}},\ \bibinfo {pages} {230602} (\bibinfo {year}
  {2012})}\BibitemShut {NoStop}%
\bibitem [{\citenamefont {Khaykovich}(2015)}]{Khaykovich2015}%
  \BibitemOpen
  \bibfield  {author} {\bibinfo {author} {\bibfnamefont {L.}~\bibnamefont
  {Khaykovich}},\ }\href@noop {} {} (\bibinfo {year} {2015}),\ \bibinfo {note}
  {private communication}\BibitemShut {NoStop}%
\end{thebibliography}

\end{document}